\documentclass[a4paper,11pt]{article}

\usepackage{bbold}
\usepackage{jheppub}
\usepackage[dvipsnames]{xcolor}
\usepackage{multirow}
\usepackage{physics}
\usepackage{tabulary}
\usepackage{listings}

\allowdisplaybreaks[4]

\graphicspath{{figures/}}

\newcommand\scalemath[2]{\scalebox{#1}{\mbox{\ensuremath{\displaystyle #2}}}}

\newcommand{\nn}{\nonumber \\}

\newcommand{\be}{\begin{equation}}
\newcommand{\ee}{\end{equation}}

\newcommand{\bnq}{\begin{equation}}
\newcommand{\enq}{\end{equation}}
\newenvironment{eqaligned}{\equation\aligned}{\endaligned\endequation}
\newcommand{\bns}{\begin{eqaligned}}
\newcommand{\ens}{\end{eqaligned}}

\newcommand{\ep}{\ensuremath{\varepsilon}}
\newcommand{\dm}{\ensuremath{\textrm{d}}}
\newcommand{\emax}{E_{\rm max}}
\newcommand{\Nep}{\ensuremath{\mathcal{N}_{\ep}}}
\newcommand{\normA}{\ensuremath{\mathcal{N}_A}}

\newcommand{\vij}{\ensuremath{v_{ij}}}

\newcommand{\intSS}{\ensuremath{\mathcal{S}{\hspace{-5pt}}\mathcal{S}}}

\newcommand{\AIzz}[3]{\ensuremath{I_{#1,#2}^{(0)}\left[#3\right]}}
\newcommand{\AIm}[1]{\ensuremath{I_{#1}^{(1)}}}

\newcommand{\AImm}[2]{\ensuremath{I_{#1,#2}^{(2)}}}
\newcommand{\MIang}[1]{\left\langle #1 \right\rangle}

\newcommand{\xa}{\ensuremath{\mathfrak{m}}}
\newcommand{\yb}{\ensuremath{\mathfrak{n}}}

\renewcommand{\arraystretch}{1.5}

\newcolumntype{t}{>{\ttfamily}l}

\preprint{
\begin{flushright}
TTP25-031,
P3H-25-067
\end{flushright}
}

\title{Integral of the  double-emission eikonal function for  two  massive emitters at an arbitrary angle}
\author{Ming-Ming Long,}
\author{Kirill Melnikov,}
\author{Andrey Pikelner}
\emailAdd{ming-ming.long@kit.edu}
\emailAdd{kirill.melnikov@kit.edu}
\emailAdd{andrey.pikelner@kit.edu}
\affiliation{Institute for Theoretical Particle Physics, KIT, Wolfgang-Gaede-Straße 1, 76131, Karlsruhe, Germany}

\abstract{ 
We present a semi-analytic  calculation of the integrated double-emission eikonal
function of two massive emitters whose momenta are at an arbitrary
angle to each other. This result is needed for extending the nested soft-collinear subtraction scheme \cite{caola:2017dug} to processes with massive partons.
 }


\allowdisplaybreaks

\begin{document}
\maketitle

\section{Introduction}
\label{sec:intro}

Subtraction schemes for higher-order QCD computations are essential for improving the 
reliability of theoretical predictions in the context of collider physics.  For next-to-next-to-leading order (NNLO) computations, significant progress in  developing such schemes for \emph{massless} partons  occurred in recent years \cite{Gehrmann-DeRidder:2005btv,caola:2017dug,currie:2013vh,czakon:2010td,czakon:2011ve,Czakon:2014oma,Cacciari:2015jma,bertolotti:2022aih,devoto:2023rpv,DelDuca:2016csb,DelDuca:2016ily,Devoto:2025kin,Devoto:2025jql,Fox:2024bfp,Gehrmann:2023dxm}. 
To extend these  results to \emph{massive} partons, further  scheme-dependent 
calculations  are required. 
In this paper, we describe a computation   of the integral of the double-emission eikonal function for two massive emitters whose momenta are at an arbitrary angle to each other.  This integral  is an important ingredient  needed for making the nested soft-collinear subtraction scheme applicable to processes with heavy quarks.   Calculations of a similar integral in the context of two different slicing schemes have been reported 
earlier in Refs.~\cite{Catani:2023tby,Liu:2024hfa}.

In principle, one can attempt to integrate the eikonal function for two massive emitters analytically,\footnote{
In fact, this is what was  done in Ref.~\cite{Liu:2024hfa} for a particular scheme choice.
We also note that the  analytic calculation for   the back-to-back kinematics in the context of the nested subtraction scheme was performed in Ref.~\cite{Bizon:2020tzr}. Earlier, a numerical computation 
of a similar quantity was performed in Ref.~\cite{Angeles-Martinez:2018mqh}.
}  but   this approach 
rapidly becomes unnecessarily complicated. Because of this, in the  current  paper we pursue a \emph{semi-analytic} approach where 
we extract  all $1/\ep$ divergences 
of the integrated eikonal function analytically, and construct a   representation for the finite remainder  that 
can be computed numerically without a regulator. 

Our approach to this problem is based on the observation  that   soft and collinear singularities of  the eikonal function 
can be easily subtracted. This observation  
was used in the calculation of the $N$-jettiness soft function at NNLO QCD described in Ref.~\cite{agarwal:2024gws}. It was also used very recently in the computation of the integral of the eikonal function  of massless and  massive emitters with momenta at arbitrary angles to each other \cite{Horstmann:2025hjp}.  
In this paper, we largely follow the 
approach of Ref.~\cite{Horstmann:2025hjp}, although 
there are important differences  between the massive-massless and the massive-massive cases.  These differences can be summarized by noticing  that the massive-massless case has stronger infra-red singularities but simpler integrals, whereas in the massive-massive case 
the situation is reversed. 

The rest of the paper is organized as follows. In Section~\ref{sec:conv} we introduce 
the eikonal function, 
explain  our conventions, and define 
the  integral of the eikonal function that is needed in the context of the nested soft-collinear subtraction scheme \cite{caola:2017dug}.  In Section~\ref{sec:singleemission},  the integral of  the single-emission eikonal function, and the so-called 
iterative contribution to the 
double-emission eikonal function are discussed. 
In Section~\ref{sec:doubleemission}
we explain how the integral of 
the non-iterative piece 
of the double-emission eikonal function 
is computed.  We describe in detail the subtraction of infra-red singularities, and explain how to express the different contributions 
that arise along the way  through  easier-to-compute    phase-space integrals. We explain 
how to extract the infra-red and collinear 
singularities from such integrals,  and construct finite remainders that are  computed  numerically. 
In Section~\ref{sect:sec5}, 
we discuss the results including the  $1/\ep$ terms, 
the implementation of the finite remainders  in a numerical code, and multiple checks that have been performed to ensure their correctness. 
We conclude in Section~\ref{sec:concl}.
Useful technical details including definitions of integrals, and aspects of  their calculation  can be found in appendices.

\section{Conventions}
\label{sec:conv}

We follow Ref.~\cite{Horstmann:2025hjp} and study   a generic partonic process
\begin{equation}
    0 \to h_1(p_1) + \cdots + h_n(p_n) + H_{n+1}(p_{n+1}) + \cdots + H_{N}(p_{N}) + f_1(k_1) + f_2(k_2),
    \label{eq2.1}
\end{equation}
where $h_{i}$ and $H_{i}$ are massless and massive partons, respectively,  and $f_{1,2}$ are
two massless, potentially unresolved partons which can be either two gluons or a
$q \bar q$ pair. We consider the double-soft limit, $k_1, k_2 \to 0$, with
all other momenta in Eq.~(\ref{eq2.1}) fixed. In this limit, the amplitude
squared of the process in Eq.~(\ref{eq2.1}) factorizes. It becomes 
\cite{Catani:1999ss}:
\begin{itemize}
\item if $f_{1,2}$ are gluons, 
\begin{equation}
\begin{split}
  \label{eq: double_soft_gg}
 \lim_{k_1,k_2 \to 0} | \mathcal{M}^{gg}(\{p\},k_1,k_2) |^2
  \approx{}
  g_{s,b}^{4}
  \bigg\{ &
            \frac{1}{2} \sum_{i,j,k,l}^{N}
            \mathcal{S}_{ij}(k_1) \mathcal{S}_{kl}(k_2)
            | \mathcal{M}^{\{(ij),(kl)\}}(\{p\}) |^2
            \\
          &
            - C_A \sum_{i,j}^{N} \mathcal{S}_{ij}(k_1,k_2) | \mathcal{M}^{(ij)}(\{p\})|^2
            \bigg\} \ ,
\end{split}
\end{equation}

\item if $f_1 = q$ and $f_2 = \bar q$, 
\begin{align}
  \label{eq: double_soft_qqb}
\lim_{k_1,k_2 \to 0}  | \mathcal{M}^{q\bar{q}}(\{p\},k_1,k_2) |^2
  \approx{}
  g_{s,b}^{4} \ T_R \sum_{i,j}^{N} \mathcal{I}_{ij}(k_1,k_2)
  | \mathcal{M}^{(ij)}(\{p\})|^2  \ .
\end{align}
\end{itemize}

Quantities that appear in the above equations include two Casimir operators of
the $SU(3)$ group, $C_A = 3, T_R = 1/2$, the bare strong coupling constant
$g_{s,b}$, as well as the color-correlated matrix elements
of the process without two soft partons
\begin{align}
  \label{eq: color_matrix}
  | \mathcal{M}^{\{(ij),(kl)\}}(\{p\})|^2
  &={}
  \langle \mathcal{M}(\{p\}) |
  \{ \boldsymbol{T}_i \cdot \boldsymbol{T}_j ,\boldsymbol{T}_k \cdot \boldsymbol{T}_l\}
  | \mathcal{M}(\{p\}) \rangle \ , \\
  | \mathcal{M}^{\{(ij)\}}(\{p\})|^2
  &={}
  \langle \mathcal{M}(\{p\}) |
  \boldsymbol{T}_i \cdot \boldsymbol{T}_j
  | \mathcal{M}(\{p\}) \rangle \ .
\end{align}
The quantities $\boldsymbol{T}_i$ are the color-charge operators \cite{Catani:1996jh}, and
$\{..,..\}$ denotes an anti-commutator. Sums in Eqs. (\ref{eq:
double_soft_gg}, \ref{eq: double_soft_qqb}) run over all pairs of hard
color-charged emitters.

In Eq.~\eqref{eq: double_soft_gg}, the term containing the product of two single-eikonal factors 
\begin{equation}
\label{eq: eikonal_g}
    \mathcal{S}_{ij}(k) = \frac{(p_i \cdot p_j)}{(p_i \cdot k)(p_j \cdot k)} \; ,
\end{equation}
is the \textit{Abelian} contribution. We note that $S_{ij}(k)$ also appears in
the single-emission eikonal function  relevant for computations at
next-to-leading order.

The \textit{non-Abelian} term, proportional to the color factor $C_A$, is more
complicated. The eikonal function $\mathcal{S}_{ij}(k_1,k_2)$ reads
\begin{align}
  \label{eq: eikonal_gg}
  \mathcal{S}_{ij}(k_1,k_2)
  ={}&
       \mathcal{S}^{0}_{ij}(k_1,k_2) +
       \left[
       m_i^2 \mathcal{S}^{m}_{ij}(k_1,k_2)
       +
       m_j^2 \mathcal{S}^{m}_{ji}(k_1,k_2)
       \right] ,
\end{align}
where quantities that appear inside the  square brackets, explicitly depend on the masses of the two
emitters, $m_{i,j}$. In addition to this explicit dependence, both functions
$\mathcal{S}^{0}_{ij}(k_1,k_2)$ and $\mathcal{S}^{m}_{ij}(k_1,k_2)$ \emph{implicitly}
depend on these masses,  since  the momenta of hard emitters are on-shell,   $p_{i,j}^2 = m_{i,j}^2$.

The first term in
Eq.~\eqref{eq: eikonal_gg}, $\mathcal{S}^{0}_{ij}(k_1,k_2)$ is the same for massless and massive emitters
\cite{Catani:1999ss}. It reads
\begin{equation}
  \begin{split}
    \label{eq: double_eikonal_s0}
    \mathcal{S}^0_{ij}(k_1,k_2) 
    ={}&   \frac{(1-\ep)}{(k_1\cdot k_2)^2} \frac{\left[(p_i\cdot k_1)(p_j\cdot k_2) + i\leftrightarrow j \right]}{(p_i\cdot k_{12})(p_j\cdot k_{12})}  \\
       & - \frac{(p_i\cdot p_j)^2}{2(p_i\cdot k_1)(p_j\cdot k_2)(p_i\cdot k_2)(p_j\cdot k_1)} \bigg[ 2 - \frac{\left[(p_i\cdot k_1)(p_j\cdot k_2) + i\leftrightarrow j \right]}{(p_i\cdot k_{12})(p_j\cdot k_{12})} \bigg] \\
       & + \frac{(p_i\cdot p_j)}{2(k_1\cdot k_2) } \bigg[ \frac{2}{(p_i\cdot k_1)(p_j\cdot k_2)} + \frac{2}{(p_j\cdot k_1)(p_i\cdot k_2)} - \frac1{(p_i\cdot k_{12})(p_j\cdot k_{12})}  \\
       & ~~ \times \left( 4 + \frac{\left[(p_i\cdot k_1)(p_j\cdot k_2) + i\leftrightarrow j \right]^2}{(p_i\cdot k_1)(p_j\cdot k_2)(p_i\cdot k_2)(p_j\cdot k_1)} \right) \bigg] \ ,
  \end{split}
\end{equation}
where we have used the abbreviation $k_{12} = k_1+k_2$. The other two
contributions in Eq.~\eqref{eq: eikonal_gg} are only relevant for the massive
emitters. The function $\mathcal{S}^{m}_{ij}(k_1,k_2)$ is given
by~\cite{Catani:2019nqv}\footnote{ A different expression
for $S_{ij}^{m}$ is found in Ref.~\cite{czakon:2011ve}. However, both
expressions give the same result after summing over $i,j$ in Eqs.~(\ref{eq:
double_soft_gg}, \ref{eq: double_soft_qqb}) thanks to colour conservation. }
\begin{equation}
  \begin{split}
    \label{eq: double_eikonal_sm}
    \mathcal{S}^{m}_{ij}(k_1,k_2) 
    ={}& \frac{(p_i\cdot p_j) (p_j \cdot k_{12})}{2(p_i\cdot k_1)(p_j\cdot k_2)(p_i\cdot k_2)(p_j\cdot k_1)(p_i \cdot k_{12})} \\
       & - \frac1{2(k_1\cdot k_2)(p_i\cdot k_{12})(p_j\cdot k_{12})} \left( \frac{(p_j\cdot k_1)^2}{(p_i\cdot k_1)(p_j\cdot k_2)} + \frac{(p_j\cdot k_2)^2}{(p_i\cdot k_2)(p_j\cdot k_1)} \right).
  \end{split}
\end{equation}

In the quark-antiquark  case, the eikonal function $\mathcal{I}_{ij}(k_1,k_2)$ reads
\begin{align}
  \label{eq: eikonal_qqb}
  \mathcal{I}_{ij}(k_1,k_2) = \frac{\left[ \left(p_i \cdot k_1 \right) \left(p_j \cdot k_2 \right) + i\leftrightarrow j \right] -\left(p_i \cdot p_j \right)\left(k_1 \cdot k_2 \right) }{\left(k_1 \cdot k_2 \right)^2\left(p_i \cdot k_{12} \right) \left( p_j \cdot k_{12} \right)} \ ,
\end{align}
and there is no difference between massive and massless emitters. 

It is  convenient to make use of the color conservation
\begin{equation}
  \sum \limits_{i=1}^{N} \boldsymbol{T}_i |{\cal M}(\{p\}) \rangle  = 0,
\end{equation}
and the symmetry of functions $S_{ij} = S_{ji}$ and $I_{ij} = I_{ji}$ to write
\begin{align}
  \sum_{i,j}^{N} \mathcal{S}_{ij}(k_1,k_2) | \mathcal{M}^{(ij)}(\{p\})|^2 &= \sum_{i<j}^{N} \widetilde{\mathcal{S}}_{ij}(k_1,k_2) | \mathcal{M}^{(ij)}(\{p\})|^2, \\
  \sum_{i,j}^{N} \mathcal{I}_{ij}(k_1,k_2) | \mathcal{M}^{(ij)}(\{p\})|^2 &= \sum_{i<j}^{N} \widetilde{\mathcal{I}}_{ij}(k_1,k_2) | \mathcal{M}^{(ij)}(\{p\})|^2,
\end{align}
where 
\begin{align}
  \label{eq:tildedEikDef}
  \widetilde{\mathcal{S}}_{ij} & = 2\mathcal{S}_{ij}  - \mathcal{S}_{ii} - \mathcal{S}_{jj},
 \\
  \widetilde{\mathcal{I}}_{ij} & = 2\mathcal{I}_{ij}  - \mathcal{I}_{ii} - \mathcal{I}_{jj}.
\end{align}

To compute the required double-soft contributions, we have to integrate the
corresponding eikonal functions $\widetilde S_{ij}$ and $\widetilde I_{ij}$ over the
phase space of two unresolved partons with momenta $k_{1,2}$. Working within the
nested soft-collinear subtraction scheme \cite{caola:2017dug}, we have to fix the reference frame, and
restrict energies of unresolved partons by introducing an upper cut-off $\emax$.
Furthermore, energies of unresolved partons must be ordered. We call the parton
with the larger (smaller) energy $\xa(\yb)$, and refer to their momenta as
$k_{\xa,\yb}$, instead of $k_{1,2}$, which describe momenta without energy
ordering.

We define the  required double-emission phase-space integrals as \cite{caola:2017dug}
\begin{align}
  \label{eq:ssintDef1}
  \intSS\left[ \mathcal{S}_{ij} \mathcal{S}_{kl} \right] &= \int [\dm k_{\xa}] [\dm k_{\yb}]
  \theta\left(\emax - k_{\xa}^0  \right)
  \theta\left(k_{\xa}^0 - k_{\yb}^0  \right)
  \mathcal{S}_{ij}(k_{\xa}) \mathcal{S}_{kl}(k_{\yb}), \\ \label{eq:ssintDef2}
  \intSS\left[ \Xi_{ij} \right] &= \int [\dm k_{\xa}] [\dm k_{\yb}]
  \theta\left(\emax - k_{\xa}^0  \right)
  \theta\left(k_{\xa}^0 - k_{\yb}^0  \right)
  \Xi_{ij}\left(k_{\xa},k_{\yb}\right), 
\end{align}
where the eikonal function $\Xi_{ij}$ is either $\widetilde{\mathcal{S}}_{ij}$ (for
$gg$ emission) or $\widetilde{\mathcal{I}}_{ij}$ (for $q \bar q$ emission), and 
\begin{equation}
    [\dm k] = \frac{\dm^{d-1}{k}}{2 k^0 (2\pi)^{d-1} },
\end{equation}
is the phase-space element.  We note that  $d=4-2\ep$ is the space-time dimension.\footnote{We use dimensional regularization to regulate soft and
collinear divergences throughout this paper.}

We  use the homogeneity of the eikonal functions to extract the dependence of the result 
on $\emax$; we  explained how to do that in Ref.~\cite{Horstmann:2025hjp}. 
Without repeating this discussion here, 
we simply quote the result for the correlated part 
\begin{equation}
  \begin{split}
    \label{eq:ssintNoThKtoL}
    \intSS\left[ \Xi_{ij} \right] 
                                  & = - \frac{1}{4\ep \emax^{4\ep}}
                                    \int [\dm l_{\xa}] [\dm l_{\yb}]
                                    \delta\left(1 - l_{\xa}\cdot P\right)
                                    \theta\left(l_{\xa}\cdot P - l_{\yb}\cdot P  \right)
                                    \Xi_{ij}\left(l_{\xa},l_{\yb}\right).
  \end{split}
\end{equation}
The  auxiliary four-vector $P$ in the above equation reads 
$P=(1,\vec{0})$.  An identical  formula applies to the product of two  single eikonal functions $\intSS[S_{ij} S_{kl}]$.
\\

Our goal is to compute the required soft integrals for 
two massive emitters. 
As we already noted,   
the masses of partons $i$ and $j$ are $m_{i,j}$, respectively. 
The squares of their  four-momenta $p_{i,j}$ are then 
$p_{i,j}^2 = m_{i,j}^2$.
We can choose a reference frame to integrate the eikonal function. The integral in Eq.~(\ref{eq:ssintNoThKtoL}) is boost-invariant, but a particular choice $P = (1,\vec 0)$ defines the  laboratory frame where both heavy partons $i$ and $j$ move with different velocities. 
In the lab frame, the momenta $p_{i,j}$ are characterized by energies or, equivalently, velocities $\beta_{i,j}$  and their directions.  We write 
\begin{equation}
p_{i,j} = m_{i,j} \gamma_{i,j}
\left ( 1, \beta_{i,j} \;  \vec \kappa_{i,j} \right),
\end{equation}
where $\gamma_{i,j} = 1/\sqrt{1-\beta_{i,j}^2}$.

However, as we discussed in Ref.~\cite{Horstmann:2025hjp}, it is beneficial to work in a different frame where one of the heavy partons is at rest.  In this frame  
\begin{equation}
P = \gamma_t \left ( 
1, v_t \vec n_t \right ),
\;\;\; p_i = m_i(1,\vec 0),
\;\;\; p_j = E_j 
\left ( 1 , v_{ij} \vec n_j 
\right ),
\end{equation}
with  $\gamma_t = \gamma_i$, $v_t =  \beta_i$ and $\vec n_t = -\vec \kappa_i$. Furthermore,  $v_{ij}$ is  the relative velocity of partons $i$ and $j$, and 
$E_j$
and $\vec n_j$ are the energy and  the direction of a parton $j$ in the rest frame of $i$.

The relation between these and the laboratory-frame  quantities is easy to establish. We find 
\begin{equation}
\begin{split}
& E_j = 
\frac{p_i p_j}{m_i }
= m_j \gamma_i \gamma_j 
\left ( 1- \beta_i \beta_j \cos \theta_{ij} \right ),
\\
& v_{ij} = \sqrt{1 - \frac{(1-\beta_i^2) (1-\beta_j^2)}{\left ( 1-\beta_i \beta_j \cos \theta_{ij} \right )^2}
},
\\
& \vec n_t \cdot \vec n_j
= \frac{1}{v_{ij}}
\left ( 1 - \frac{1-\beta_i^2}{1-\beta_i \beta_j \cos \theta_{ij}}
\right ),
\end{split}
\label{eq2.22}
\end{equation}
where $\cos \theta_{ij} = \vec \kappa_i \cdot \vec \kappa_j$. As we will see, the integrated eikonal function depends on  $\vec n_t \cdot \vec n_j$, $v_{ij}$, and $v_t$,   and all these quantities can be expressed in terms of the parameters in the laboratory  frame using  Eq.~(\ref{eq2.22}).

\section{Single emission and iterations} 
\label{sec:singleemission}

To compute the iterated part of the double-real emission function, we need to 
integrate the product of two single-emission eikonal functions. The integral reads 
\be
\intSS[S_{ij} S_{km}] 
= -\frac{1}{4 \ep E_{\rm max}^{4 \ep}} \int [{\rm d} l_\xa] [{\rm d} l_\yb] 
\delta(1-  l_\xa \cdot P ) 
\theta( l_\xa \cdot  P - l_\yb \cdot  P) \;
S_{ij}(l_\xa) \; S_{km}(l_\yb).
\label{eq3.1}
\ee
The gluon momenta $ l_{\xa,\yb}$ are defined as $l_{\xa,\yb} = l_{\xa,\yb}^{0}(1,\vec n_{\xa, \yb})$, 
with $\vec n_{\xa,\yb}^2 = 1$.
We calculate  the integral in 
Eq.~(\ref{eq3.1})
 directly in the laboratory frame, i.e. $P = (1, \vec 0)$.
Writing $l^0_\yb = \omega  \; l_\xa^0$  and integrating over $\omega$,  
we find 
\be
\intSS[S_{ij} S_{km}] 
= \frac{N_\ep^2}{8 \ep^2 E_{\rm max}^{4 \ep}} \left \langle \frac{\rho_{ij}}{\rho_{i \xa} \rho_{j \xa}}
\right \rangle_\xa \; 
\left \langle \frac{\rho_{km}}{\rho_{k \yb} \rho_{m \yb}}
\right \rangle_\yb,
\label{eq3.2}
\ee
where $\rho_{xy} = 1 - 
\vec \beta_x \cdot \vec \beta_y$, with $\vec \beta_{\xa,\yb} = 
\vec n_{\xa, \yb}$.
Furthermore, we use 
\be
N_\ep = \frac{\Omega^{(d-1)}}{2 (2\pi)^{d-1}},
\label{eq3.3}
\ee
and 
\be
\Big \langle \dots \Big \rangle_x = 
\int \frac{{\rm d} \Omega^{(d-1)}_x}{\Omega^{(d-1)}} \cdots,
\ee
is the integral over directions of the parton $x$.

Using integrals defined in Appendix~\ref{app: two_mass_int}, we write Eq.~(\ref{eq3.2}) as
\be
\intSS[S_{ij} S_{km}] 
= \frac{N_\ep^2}{8 \ep^2 E_{\rm max}^{4 \ep}} \; I_{11}^{(2)}[\rho_{ii},\rho_{jj}, \rho_{ij} ] \; 
I_{11}^{(2)}\left [\rho_{kk},\rho_{mm}, 
\rho_{km} \right ].
\label{eq3.5}
\ee

Integrals $I^{(2)}_{11}$ 
cannot be computed in closed form for arbitrary $d$ and require an expansion in $\ep$.
The depth of the required  expansion depends on 
whether lines $i,j,k,m$ are massive or massless. If all lines are massive, the integrals are finite; hence, each of them needs to be expanded to ${\cal O}(\ep^2)$. If, on the other hand, there are  one or more massless partons, additional singularity is generated
by the angular integration, 
and 
the angular integral for massive lines through 
${\cal O}(\ep^3)$ is needed. 

We note in passing that these very deep expansions  may not be necessary. Indeed,   for the purpose of subtractions, the iterative piece 
will have to be combined with an iteration of Catani's $I^{(1)}$ operator \cite{Catani:1996jh,Catani:2002hc},  appearing in the virtual corrections; for the massless case, this mechanism  was explained in Ref.~\cite{devoto:2023rpv}.
Similarly, also  for the massive case it  should lead to 
a cancellation of the highest 
$1/\ep^2$  singularities, without destroying the factorized form as in Eq.~(\ref{eq3.5}). Because of this, the required depth of the $\ep$-expansion will  be reduced.

\section{Non-iterative double-real emission contribution}
\label{sec:doubleemission}

We consider the integral of the correlated contribution to 
the double-real eikonal function,  defined  
in Eq.~(\ref{eq:ssintNoThKtoL}). We will compute it in the rest frame of the parton $i$. Using the expression for $P$ and other momenta in the rest frame 
of $i$  given in Section~\ref{sec:conv},
and integrating over the energy of the gluon $\xa$, we find
\be
\intSS[\Xi_{ij}] = \frac{\mathcal{N}_A}{\ep}
\left\langle
\int_0^{\infty} \frac{\dm \omega}{\omega^{1+2\ep}} \rho_{t\xa}^{4\ep} \theta(\rho_{t\xa}-\omega \rho_{t\yb})
\left[\omega^2 \Xi_{ij}(\xa, \yb)\right]
\right\rangle_{\xa\yb},
\label{eq4.1}
\ee
where $\rho_{tx} = 1 - v_t \vec n_t \cdot \vec n_x$, $x = \xa, \yb$, 
\bnq
\mathcal{N}_A = -\frac{\Nep^2}{4} \left(\frac{\emax}{\gamma_t}\right)^{-4\ep},
\enq
and the momenta of gluons $\xa$ and $\yb$ to be used in 
Eq.~(\ref{eq4.1}) are 
$l_\xa = (1, \vec n_\xa)$ and $l_\yb = \omega (1, \vec n_\yb)$.

Eq.~(\ref{eq4.1}) is a convenient starting point for  the  computation. We will focus on the calculation of  the gluon eikonal function  since it is more general than the quark one. Hence, we identify  $\Xi_{ij}$ with $ \widetilde{S}_{ij}$, c.f.   Eq.~(\ref{eq:tildedEikDef}); we  will refer to $\intSS[\widetilde S_{ij}]$ as $G_{ij}$.
The integral in Eq.~(\ref{eq4.1}) cannot be computed  numerically
right away because of divergences. For two massive emitters, these divergences appear in just two cases -- i) when the  gluon $\yb$ becomes soft, $\omega \to 0$, and 
ii) when the gluons $\xa$ and $\yb$ become collinear to each other.  We will iteratively subtract 
these singularities from the integrand in Eq.~(\ref{eq4.1}) to construct a finite quantity that can be calculated numerically. 
However, we stress  that because of the overall 
$1/\ep$ factor in Eq.~(\ref{eq4.1}), it is highly non-trivial to compute all $1/\ep$ poles analytically even if all sources of divergences have been removed from the integrand in that equation.
 
We start with the soft subtraction and write  
\bnq
G_{ij} = {\cal S}_{\omega}[G_{ij}] + \overline{{\cal S} }_{\omega}[G_{ij}].
\enq
The first term corresponds to the strongly-ordered limit of $G_{ij}$, 
\bnq
{\cal S}_{\omega}[G_{ij}] = 
\frac{\mathcal{N}_A}{\ep}
\left\langle
\int_0^{\infty} \frac{\dm \omega}{\omega^{1+2\ep}} \; \rho_{t\xa}^{4\ep} \; \theta(\rho_{t\xa}-\omega \rho_{t\yb}) \;
S_\omega \left[\omega^2 \widetilde{S}_{ij}(\xa, \yb)\right]
\right\rangle_{\xa\yb},
\enq
where the operator $S_\omega$ extracts the leading ${\cal O}(1/\omega^2)$ singularity from the eikonal function $\widetilde S_{ij}$. 
We will discuss the computation of this quantity in the next section. 

The second term  
\bnq
\overline{{\cal S}}_{\omega}[G_{ij}] = 
\frac{\normA}{\ep}
\left\langle
\int_0^{\infty} \frac{\dm \omega}{\omega^{1+2\ep}} \rho_{t\xa}^{4\ep} \theta(\rho_{t\xa}-\omega \rho_{t\yb})
\overline{S}_{\omega}\left[\omega^2 \widetilde{S}_{ij}(\xa, \yb)\right]
\right\rangle_{\xa\yb},
\enq
involves the  operator $\overline {S}_\omega = 1- S_\omega$. Hence, 
it does not possess a  soft singularity since it is explicitly  subtracted. It remains to isolate and remove the collinear $ \xa || \yb$ singularity from it.  While such a subtraction is 
straightforward, we need to do it in such a way, that all divergent contributions can be computed analytically. 

To this end, we found it convenient to write $\overline{{\cal S}}_{\omega}[G_{ij}]$ as the sum of two terms 
\bnq
\overline{{\cal S}}_{\omega}[G_{ij}] = 
 \overline{{\cal S}}_{\omega}[G^{(0)}_{ij}]
+
\overline{{\cal S}}_{\omega}[\Delta G_{ij}].
\enq
The first term reads  
\bnq
\overline{{\cal S}}_{\omega}[G^{(0)}_{ij}]
=
\frac{\normA}{\ep}
\left\langle
\int_0^{\infty} \frac{\dm \omega}{\omega^{1+2\ep}} \; \theta(1-\omega) \; \overline{S}_{\omega}
\left[\omega^2 \widetilde{S}_{ij}(\xa, \yb)\right]
\right\rangle_{\xa\yb}.
\label{eq4.8}
\enq
It  corresponds to  the soft-subtracted 
integral in case when the  laboratory frame and the rest frame of the 
parton $i$ coincide. The function $\widetilde S_{ij}$ in this case depends on a single direction, which makes the integration much  simpler. 

The second  quantity
$\overline{{\cal S}}_{\omega}[\Delta G_{ij}]$ 
reads 
\bnq
\overline{\cal S}_{\omega}[\Delta G_{ij}] =
\frac{\normA}{\ep}
\left\langle
\int_0^{\infty} \frac{\dm \omega}{\omega^{1+2\ep}} \;
f(\omega, \rho_{t \xa},\rho_{t \yb} ) \;
\overline{S}_{\omega}\left[\omega^2 \widetilde{S}_{ij}(\xa, \yb)\right]
\right\rangle_{\xa\yb},
\label{eq4.9}
\enq
where 
\be
f(\omega, \rho_{t \xa}, 
\rho_{t \yb} ) = 
    \rho_{t\xa}^{4\ep} \theta(\rho_{t\xa}-\omega \rho_{t\yb})
    -\theta(1-\omega).
\ee
The  integral in Eq.~(\ref{eq4.9}) is divergent  because of the collinear  $\xa || \yb$ singularity of the integrand.  
However, as we will see shortly 
this divergence is softer than the divergence of the  
quantity $\overline{\cal S}_\omega[G_{ij}]$.

To subtract the $\xa || \yb$ divergence, 
we write  
$\overline {S}_\omega[\Delta G_{ij}]$ as the sum of two terms 
\bnq
\overline{{\cal S}}_{\omega}[\Delta G_{ij}] = 
\overline{{\cal S}}_{\omega}[\Delta G_{ij}]_{\xa || \yb} + \overline{{\cal S}}_{\omega}[\Delta G_{ij}]_{\rm fin}, 
\enq
where 
\be 
\begin{split}
& \overline{{\cal S}}_{\omega}[\Delta G_{ij}]_{\xa || \yb} 
= \frac{\mathcal{N}_A}{\ep}
\left\langle
\int_0^{\infty} \frac{\dm \omega}{\omega^{1+2\ep}} \; 
C_{\xa \yb} \; f(\omega,\rho_{t \xa}, 
\rho_{t \yb}) \; 
\overline{S}_{\omega} \; \left[\omega^2 \widetilde{S}_{ij}(\xa, \yb)\right]
\right\rangle_{\xa\yb},
\\
& \overline{{\cal S}}_{\omega}[\Delta G_{ij}]_{\rm fin} 
= \frac{\mathcal{N}_A}{\ep}
\left\langle
\int_0^{\infty} \frac{\dm \omega}{\omega^{1+2\ep}} 
\; 
\overline {C}_{\xa \yb} \; f(\omega,\rho_{t\xa}, 
\rho_{t \yb}) \; 
\overline{S}_{\omega}
\; \left[\omega^2 \widetilde{S}_{ij}(\xa, \yb)\right]
\right\rangle_{\xa\yb}.
\end{split}
\ee
The operator $\overline {C}_{\xa \yb}$ is  
defined as $\overline {C}_{\xa \yb} = 1 - C_{\xa \yb}$.
The  operator
$C_{\xa\yb}$ extracts the non-integrable part of the $\xa || \yb$  collinear  limit of the integrand, but it does not act on the angular phase space. Since 
\bnq
C_{\xa\yb} f(\omega,\rho_{t \xa},
\rho_{t \yb})
=(\rho_{t\xa}^{4\ep}-1) \; \theta(1-\omega) \; C_{\xa\yb},
\enq
we find 
\bns
\overline{{\cal S}}_{\omega}[\Delta G_{ij}]_{\xa || \yb}  
&= \frac{\mathcal{N}_A}{\ep}
\left\langle
\int_0^{\infty} \frac{\dm \omega}{\omega^{1+2\ep}} (\rho_{t\xa}^{4\ep}-1) \; \theta(1-\omega)
\; C_{\xa\yb} \; \overline{S}_{\omega}
\left[\omega^2 \widetilde{S}_{ij}(\xa, \yb)\right]
\right\rangle_{\xa\yb}.
\ens

To simplify 
$\overline {\cal S}_\omega[\Delta G_{ij} ]_{\rm fin}$, we  split $\overline{C}_{\xa \yb} f(\omega,\rho_{t\xa},\rho_{t \yb})
$ into two terms
\bns
\overline{C}_{\xa\yb} 
\left[
    \rho_{t\xa}^{4\ep} \theta(\rho_{t\xa}-\omega \rho_{t\yb})
    -\theta(1-\omega)
\right]
&=
\overline{C}_{a} + \overline{C}_{b},
\ens
using the following decomposition $\rho_{t\xa}^{4\ep} = 1+(\rho_{t\xa}^{4\ep}-1)$. 
The two terms are defined as follows  
\bns
\overline{C}_{a} 
&=
\overline{C}_{\xa\yb} 
\left[
    \theta(\rho_{t\xa}-\omega \rho_{t\yb})
    -\theta(1-\omega)
\right]
=
\theta(\rho_{t\xa}-\omega \rho_{t\yb}) -\theta(1-\omega),\\
\overline{C}_{b} 
&=
\overline{C}_{\xa\yb} 
\left[
    (\rho_{t\xa}^{4\ep}-1)\theta(\rho_{t\xa}-\omega \rho_{t\yb})
\right]
=
(\rho_{t\xa}^{4\ep}-1)
\left[\theta(\rho_{t\xa}-\omega \rho_{t\yb})-\theta(1-\omega)C_{\xa\yb} \right].
\label{eq4.15}
\ens
We then write 
\bnq
\overline{{\cal S}}_{\omega}[\Delta G_{ij}]_{\rm fin}
=
\overline{{\cal S}}_{\omega}[\Delta G_{ij}]_{\rm fin, a}
+
\overline{{\cal S}}_{\omega}[\Delta G_{ij}]_{\rm fin, b},
\enq
where 
\bns
\overline{{\cal S}}_{\omega}[\Delta G_{ij}]_{\rm fin, a}
&=
\frac{\normA}{\ep}
\left\langle
\int_0^{\infty} \frac{\dm \omega}{\omega^{1+2\ep}} 
\left ( \theta(\rho_{t\xa} 
- \omega \rho_{ t\yb} 
) 
- \theta(1-\omega) 
\right )
\overline{S}_{\omega}\left[\omega^2 \widetilde{S}_{ij}(\xa, \yb)\right]
\right\rangle_{\xa\yb}, \\
\overline{{\cal S}}_{\omega}[\Delta G_{ij}]_{\rm fin,  b}
&=
\frac{\normA}{\ep}
\left\langle
\int_0^{\infty} \frac{\dm \omega}{\omega^{1+2\ep}} 
\overline{C}_{b}
\overline{S}_{\omega}\left[\omega^2 \widetilde{S}_{ij}(\xa, \yb)\right]
\right\rangle_{\xa\yb}.
\ens
The quantity $\overline{\cal S}_\omega[\Delta G_{ij}]_{\rm fin,b}$ does not require further manipulations since its integrand is ${\cal O}(\ep)$, c.f. Eq.~(\ref{eq4.15}); hence, we compute it numerically.  

The quantity $\overline{\cal S}_\omega[\Delta G_{ij}]_{\rm fin,a}$ is also finite, but its 
integrand is not suppressed by $\ep$. Therefore, because of the $1/\ep$ prefactor, it needs to be expanded to linear order in 
$\ep$. To facilitate such an expansion, we make use of the  following identity
\bnq
\overline{S}_{\omega}
\left[\omega^2 \widetilde{S}_{ij}(\xa, \yb)\right]
\Bigg |_{\substack{\omega \rightarrow \frac{1}{\omega}  \\ \xa \leftrightarrow \yb}}
-
\overline{S}_{\omega}
\left[\omega^2 \widetilde{S}_{ij}(\xa, \yb)\right]
= \Delta_{ij},
\enq
where
\bnq
\Delta_{ij} = 
\frac
{
  \left(\rho_{\xa j}-\rho_{\yb j}\right) \left(\rho_{\xa \yb}-\rho_{\xa j}-\rho_{\yb j}\right)
}{
  \rho_{\xa \yb} \rho_{\xa j} \rho_{\yb j}
}
+\left(1-\vij^2\right)
\frac
{
  \left(\rho_{\xa j}-\rho_{\yb j}\right) \left(\rho_{\xa j}+\rho_{\yb j}-\rho_{\xa \yb}\right)
}{
  \rho_{\xa \yb} \rho_{\xa j}^2 \rho_{\yb j}^2
}.
\label{eq4.20}
\enq
It follows that 
\bns
\overline{{\cal S}}_{\omega}[\Delta G_{ij}]_{\rm fin, a}
=
\overline{{\cal S}}_{\omega}[\Delta G_{ij}]_{\rm fin, a_1}
+ \overline{{\cal S}}_{\omega}[\Delta G_{ij}]_{\rm fin, a_2},
\ens
where 
\bnq
\overline{{\cal S}}_{\omega}[\Delta G_{ij}]_{\rm fin, a_1}
=
\frac{\normA}{2\ep}
\left\langle
  \int_0^{\infty} \frac{\dm \omega}{\omega^{1-2\ep}} 
  (\theta(\omega\rho_{t\yb}-\rho_{t\xa})-\theta(\omega-1) ) 
  \Delta_{ij}
\right\rangle_{\xa\yb},
\label{eq4.22}
\enq
and
\be
\begin{split}
\overline{{\cal S}}_{\omega}[\Delta G_{ij}]_{\rm fin, a_2}
= &
\frac{\normA}{\ep}
\left \langle
  \int_0^{\infty} \frac{\dm \omega}{\omega} \; 
   \frac{\omega^{-2\ep}-\omega^{2\ep}}{2} 
\right.    \\
& \left . \times  \left[\theta(\rho_{t\xa}-\omega \rho_{t\yb})-\theta(1-\omega) \right]
  \overline{S}_{\omega}
\left[\omega^2 \widetilde{S}_{ij}(\xa, \yb)\right]
\right\rangle_{\xa\yb}.
\end{split}
\ee
The integrand of the latter quantity is ${\cal O}(\ep)$, 
so that it can be computed numerically without further ado. 

Putting everything together, we write $G_{ij}$ as 
\bnq
G_{ij} = 
{\cal S}_{\omega}[G_{ij}] + \overline{{\cal S}}_{\omega}[G^{(0)}_{ij}]
+ \overline{{\cal S}}_{\omega}[\Delta G_{ij}]_{\xa || \yb}
+\overline{\cal S}_\omega[\Delta G_{ij}]_{\rm fin, a_1} + 
\overline{{\cal S}}_{\omega}[\Delta G_{ij}]_{\mathrm{num}},
\label{eq4.23}
\enq
where four first terms on the right-hand side require 
further work, and the last term  can be computed numerically by taking the $\ep \to 0$ limit. It reads 
\bns
& \overline{S}_{\omega}[\Delta G_{ij}]_{\mathrm{num}} 
= 
\overline{S}_{\omega}[\Delta G_{ij}]_{\rm fin, a_2} + \overline{S}_{\omega}[\Delta G_{ij}]_{\rm fin, b} 
\\
& = 
4 \normA
\left\langle
  \int_0^{\infty} \frac{\dm \omega}{\omega} \ln \rho_{t \xa} \left[\theta(\rho_{t\xa}-\omega \rho_{t\yb})-\theta(1-\omega) C_{\xa \yb}\right]
  \overline{S}_{\omega}
\left[\omega^2 \widetilde{S}_{ij}(\xa, \yb)\right]
\right\rangle_{\xa\yb} 
\\
& -2 \normA
\left\langle
  \int_0^{\infty} \frac{\dm \omega}{\omega} \ln \omega \left[\theta(\rho_{t\xa}-\omega \rho_{t\yb})-\theta(1-\omega) \right]
  \overline{S}_{\omega}
\left[\omega^2 \widetilde{S}_{ij}(\xa, \yb)\right]
\right\rangle_{\xa\yb}.
\label{eq4.24}
\ens
We note that the four terms that require  additional  work have different degrees of divergence, 
that we illustrate below
\be
\begin{split}
& {\cal S}_{\omega}[G_{ij}] \sim 
{\cal O}(\ep^{-3}), \;\;\;\; 
\overline{{\cal S}}_{\omega}[G^{(0)}_{ij}] 
\sim {\cal O}(\ep^{-2}), 
\nonumber \\
& \overline{{\cal S}}_{\omega}[\Delta G_{ij}]_{\xa || \yb}
\sim {\cal O}(\ep^{-1}),
\;\;\;\; 
\overline{\cal S}_\omega[\Delta G_{ij}]_{\rm fin, a_1} 
\sim {\cal O}(\ep^{-1}).
\end{split}
\ee
Hence, the most challenging quantity to 
deal with is the  strongly-ordered 
contribution ${\cal S}_\omega[G_{ij}]$. We discuss it in the next section.

\subsection{The integral of the   strongly-ordered eikonal function ${\cal S}_{\omega}[G_{ij}]$ }
\label{ssect4.1}

In this section, we discuss the integration of the strongly-ordered eikonal function $S_{\omega}[G_{ij}]$,
\bnq
{\cal S}_{\omega}[G_{ij}] = 
\frac{\mathcal{N}_A}{\ep}
\left\langle
\int_0^{\infty} \frac{\dm \omega}{\omega^{1+2\ep}} \rho_{t\xa}^{4\ep} \theta(\rho_{t\xa}-\omega \rho_{t\yb})
S_{\omega}\left[\omega^2 \widetilde{S}_{ij}(\xa, \yb)\right]
\right\rangle_{\xa\yb}.
\enq
For the two-gluon case, we find 
\bns
S_{\omega}\left[\omega^2 \widetilde{S}_{ij}(\xa, \yb)\right] &= 
\frac{2}{\rho _{\xa j} \rho _{\xa \yb}}
+\frac{2}{\rho _{\xa \yb} \rho _{\yb j}}
-\frac{2}{\rho _{\xa j} \rho _{\yb j}}
-\frac{1}{\rho _{\xa \yb}}
+\frac{1}{\rho _{\yb j}}
-\frac{\rho _{\xa j}}{\rho _{\xa \yb} \rho _{\yb j}}
\\
&
\quad 
+(1-\vij^2)
\left(
\frac{1}{\rho _{\xa j}^2 \rho _{\yb j}}
-\frac{1}{\rho _{\xa j}^2 \rho _{\xa \yb}}
-\frac{1}{\rho _{\xa j} \rho _{\xa \yb} \rho _{\yb j}}
\right).
\ens
The integration over $\omega$ is straightforward
\bnq
\int_0^{\infty} \frac{\dm \omega}{\omega^{1+2\ep}} \rho_{t\xa}^{4\ep} \theta(\rho_{t\xa}-\omega \rho_{t\yb})
=
-\frac{1}{2\ep}\rho_{t\xa}^{2\ep}\rho_{t\yb}^{2\ep},
\enq
and we obtain 
\bns
{\cal S}_{\omega}[G_{ij}] = 
-\frac{\mathcal{N}_A}{2\ep^2}
\Bigg\langle
\rho_{t\xa}^{2\ep}\rho_{t\yb}^{2\ep}
&
\left[
\frac{2}{\rho _{\xa j} \rho _{\xa \yb}}
+\frac{2}{\rho _{\xa \yb} \rho _{\yb j}}
-\frac{2}{\rho _{\xa j} \rho _{\yb j}}
-\frac{1}{\rho _{\xa \yb}}
+\frac{1}{\rho _{\yb j}}
-\frac{\rho _{\xa j}}{\rho _{\xa \yb} \rho _{\yb j}}
\right. \\
&
\left.
+(1-\vij^2)
\left(
\frac{1}{\rho _{\xa j}^2 \rho _{\yb j}}
-\frac{1}{\rho _{\xa j}^2 \rho _{\xa \yb}}
-\frac{1}{\rho _{\xa j} \rho _{\xa \yb} \rho _{\yb j}}
\right)
\right]
\Bigg\rangle_{\xa\yb}.
\ens
This expression involves  terms with and without $\rho_{\xa \yb}$, 
and we find it convenient to separate them. Hence, we write 
\be
{\cal S}_\omega[G_{ij}]
= {\cal S}_{\omega}[G_{ij}]_{\text{F}}
+
{\cal S}_{\omega}[G_{ij}]_{\text{NF}}.
\ee
The first term does not contain $1/\rho_{\xa \yb}$; it reads 
\bns
{\cal S}_{\omega}[G_{ij}]_{\text{F}} = 
-\frac{\mathcal{N}_A}{2\ep^2}
\Bigg\langle
\frac{ \rho_{t\yb}^{2\ep} }{\rho_{\yb j}}
\Bigg \rangle_\yb 
\Bigg \langle \; \rho_{t\xa}^{2\ep}
\left[ 1
-\frac{2}{\rho _{\xa j} }
+\frac{1-\vij^2}{\rho _{\xa j}^2 }
\right]
\Bigg\rangle_{\xa}.
\ens
The angular  integrations are actually finite, 
so that ${\cal S}_\omega[G_{ij}]_{\rm F} \sim \ep^{-2}$.
Using integrals defined in the 
appendix, we derive
\bnq
S_{\omega}[G_{ij}]_{\text{F}} = 
-\frac{\mathcal{N}_A}{2\ep^2} \; 
\AImm{-2\ep}{1}
\left[
	\AIm{-2\ep}\left[\rho_{tt} \right]
	-2\AImm{-2\ep}{1}
	+(1-\vij^2)\AImm{-2\ep}{2}
\right].
\label{eq4.31}
\enq
We note that unless their arguments are shown  explicitly,  all integrals in the above 
formula should be interpreted as $I[\rho_{tt},\rho_{jj}, \rho_{tj}]$.\footnote{To avoid 
confusion, we note that in this section we always work in the rest frame of the parton $i$. This implies that 
$\rho_{tt} = 1-v_t^2$, 
$\rho_{jj} = 1-\vij^2, 
\rho_{tj} = 1-\vij  v_t  \vec n_t \cdot \vec n_j$. 
The relation of various quantities in the above equations to  their counterparts in the laboratory frame is given in  Eq.~(\ref{eq2.22}).}

The remaining, 
\emph{non-factorizable}, part is more complicated. It reads
\bns
{\cal S}_{\omega}[G_{ij}]_{\text{NF}} = 
-\frac{\mathcal{N}_A}{2\ep^2}
\Bigg\langle
\frac{
\rho_{t\xa}^{2\ep}
\rho_{t\yb}^{2\ep}
}{\rho_{\xa \yb}}
&
\left[
\frac{2}{\rho_{\xa j}} 
+\frac{2}{ \rho_{\yb j}}
-1
-\frac{\rho _{\xa j}}{ \rho _{\yb j}}
\right. \\
&
\left.
-(1-\vij^2)
\left(
\frac{1}{\rho _{\xa j}^2 }
+\frac{1}{\rho _{\xa j} \rho _{\yb j}}
\right)
\right]
\Bigg\rangle_{\xa\yb}.
\label{eq4.33}
\ens
To simplify Eq.~(\ref{eq4.33}), we use  an  opportunity  to rename  $\xa \leftrightarrow \yb$, rewriting  it in the following way
\bns
{\cal S}_{\omega}[G_{ij}]_{\text{NF}} = 
-\frac{\mathcal{N}_A}{2\ep^2}
\Bigg\langle
\frac{\rho_{t\xa}^{2\ep}\rho_{t\yb}^{2\ep}}{\rho _{\xa \yb}}
\left[
\frac{4-\rho_{\yb j}}{\rho _{\xa j} }
-1
-(1-\vij^2)
\left(
\frac{1}{\rho _{\xa j}^2}
+\frac{1}{\rho _{\xa j} \rho _{\yb j}}
\right)
\right]
\Bigg\rangle_{\xa\yb}.
\label{eq: Somega_NF}
\ens

Furthermore, we find it convenient to write  
\begin{equation}
\rho_{t\xa}^{2\ep}\rho_{t\yb}^{2\ep}
=
\frac{\rho_{t\xa}^{4\ep}+\rho_{t\yb}^{4\ep}}{2}-\frac{ \left(\rho_{t \xa}^{2\ep}-\rho_{t \yb}^{2\ep}\right)^2}{2}.
\label{eq4.34}
\end{equation}
The last term on the right-hand side 
in Eq.~(\ref{eq4.34})  provides a finite contribution to $S_\omega[G_{ij}]_{\rm NF}$, which 
can be computed numerically right away. Hence, we define 
\bns
{\cal S}_{\omega}[G_{ij}]_{\rm NF}^{\rm num} = 
\frac{\mathcal{N}_A}{4 \ep^2}
\Bigg\langle
\frac{\left(\rho_{t \xa}^{2\ep}-\rho_{t \yb}^{2\ep} \right)^2}{\rho _{\xa \yb}}
\left[
\frac{4-\rho_{\yb j}}{\rho _{\xa j} }
-1
-(1-\vij^2)
\left(
\frac{1}{\rho _{\xa j}^2}
+\frac{1}{\rho _{\xa j} \rho _{\yb j}}
\right)
\right]
\Bigg\rangle_{\xa\yb}.
\label{eq4.35}
\ens

The remaining 
contribution reads 
\bns
{\cal S}_{\omega}[G_{ij}]_{\rm NF}^{\rm div} = 
-\frac{\mathcal{N}_A}{4 \ep^2}
\Bigg\langle
\frac{\rho_{t\xa}^{4\ep}+\rho_{t \yb}^{4 \ep}
}{\rho _{\xa \yb}}
\left[
\frac{4-\rho_{\yb j}}{\rho _{\xa j} }
-1
-(1-\vij^2)
\left(
\frac{1}{\rho _{\xa j}^2}
+\frac{1}{\rho _{\xa j} \rho _{\yb j}}
\right)
\right]
\Bigg\rangle_{\xa\yb}.
\ens
Performing further $\xa \leftrightarrow  \yb$ redefinition, we  remove  $\rho_{t\yb}^{4 \ep}$
from the integrand  and obtain 
\bns
{\cal S}_{\omega}[G_{ij}]_{\rm NF}^{\rm div} = 
& -\frac{\mathcal{N}_A}{4 \ep^2}
\Bigg\langle
\frac{\rho_{t\xa}^{4\ep}}{\rho _{\xa \yb}}
\left[
\frac{4-\rho_{\yb j}}{\rho _{\xa j} }
+\frac{4-\rho_{\xa j}}{\rho _{\yb j} }
-2
\right. 
\\
& \left.  -(1-\vij^2)
\left(
\frac{1}{\rho _{\xa j}^2} + 
\frac{1}{\rho^2_{\yb j}}
+\frac{2}{\rho _{\xa j} \rho _{\yb j}}
\right)
\right]
\Bigg\rangle_{\xa\yb}.
\ens
To write  the above equation in a form convenient for further analysis,  we introduce the following  integral family 
\begin{equation}
  I^{(2)}_{a_1, a_2, a_3, a_4+b_4\ep} 
  =
\left\langle 
  \frac{1}{
  \rho_{\xa \yb}^{a_1}
  \rho_{\xa j}^{a_2}
  \rho_{\yb j}^{a_3}
  \rho_{t \xa}^{a_4+b_4 \ep}
  }
\right\rangle_{\xa \yb},
  \label{eq:}
\end{equation}
and use it to write 
\begin{equation}
  \begin{split}
  S_{\omega}[G_{ij}]_{\text{NF}}^{\rm div}
  =&
 -\frac{\normA}{4\ep^2}
  \Bigg\{
   4I^{(2)}_{1,1,0,-4\ep}
   +4I^{(2)}_{1,0,1,-4\ep}
   -I^{(2)}_{1,1,-1,-4\ep}
   -I^{(2)}_{1,-1,1,-4\ep}
   -2 I^{(2)}_{1,0,0,-4\ep}
   \\
    &
    -(1-\vij^2) \left[
   I^{(2)}_{1,2,0,-4\ep}
   +I^{(2)}_{1,0,2,-4\ep}
   +2 I^{(2)}_{1,1,1,-4\ep}
    \right]
  \Bigg\}.
  \end{split}
  \label{eq4.39}
\end{equation}
Finally, the strongly-ordered contribution  reads 
\be
{\cal S}_\omega[G_{ij}]
= {\cal S}_{\omega}[G_{ij}]_{\text{F}}
+
{\cal S}_{\omega}[G_{ij}]_{\text{NF}}^{\rm div}
+ {\cal S}_{\omega}[G_{ij}]_{\text{NF}}^{\rm num},
\label{eq4.40}
\ee
where the relevant contributions are given in Eqs.~(\ref{eq4.31},\ref{eq4.35},\ref{eq4.39}).

\subsection{The soft-subtracted, double-collinear contirbution}
As the next step, we analyse the soft-subtracted, double-collinear contribution. It reads  
\bns
\overline{{\cal S}}_{\omega}[\Delta G_{ij}]_{\xa || \yb} = 
\frac{\normA}{\ep} 
\left\langle
\int_0^{\infty} \frac{\dm \omega}{\omega^{1+2\ep}} (\rho_{t\xa}^{4\ep}-1) \theta(1-\omega)C_{\xa\yb}\overline{S}_{\omega}
\left[\omega^2 \widetilde{S}_{ij}(\xa, \yb)\right]
\right\rangle_{\xa\yb}.
\label{eq:dcIntSectors}
\ens
This 
term possesses a $1/\ep$ singularity at most,  because $\rho_{t \xa}^{4\ep} - 1 \sim {\cal O}(\ep)$.
To isolate it, 
we compute the  $\xa || \yb$ limit of the soft-subtracted 
eikonal function and find 
\begin{equation}
C_{\xa\yb} \; \overline{S}_{\omega}
\left[\omega^2 \widetilde{S}_{ij}(\xa, \yb)\right]
=
\frac{4 (\vec n_j \cdot \vec r )^2 \omega ^2 (1-\ep ) \vij^2}{(\omega +1)^4 \rho _{\xa j}^2 \rho _{\xa \yb}}
+\frac{4 \omega  \left[\left(\rho _{\xa j}-2\right) \rho _{\xa j}+1-\vij^2\right]}{(\omega +1)^2 \rho _{\xa j}^2 \rho _{\xa \yb}},
\label{eq4.42}
\end{equation}
where $\vec r$  is the unit vector  defined by the following equation 
\begin{equation}
  \label{eq:kPrepDef}
  \vec n_\yb = \cos{\theta_{\xa \yb}} \vec n_\xa + \sin{\theta_{\xa \yb}} \vec r,
  \quad
  \vec r \cdot \vec n_\xa = 0.
\end{equation}
We note that  by 
setting $v_{ij} \to  1$   in 
Eq.~(\ref{eq4.42}), we reproduce the soft-subtracted double-collinear contribution for  the massive-massless case, discussed  in Ref.~\cite{Horstmann:2025hjp}.

To proceed with the calculation of $\overline{{\cal S}}_{\omega}[\Delta G_{ij}]_{\xa || \yb}$,
we integrate over directions of a parton $\yb$ in Eq.~(\ref{eq:dcIntSectors}), 
starting with the integration over $\vec r$. Using
\begin{equation}
  r^i r^j \to 
  \frac{\delta^{ij} -\vec n^i_\xa \vec n^j_\xa}{d-2},
\end{equation}
we obtain 
\begin{align}  
  \left \langle C_{\xa \yb} \overline {S}_\omega \left [ \omega^2 \widetilde{S}_{ij}(\xa,\yb) \right ] 
  \right \rangle_{\kappa}
  =
  -2
  \frac{ \omega  (2 + 3 \omega + 2 \omega^2)}{(1+\omega)^4}
  \left[
  \frac{(2-\rho_{\xa j}) }{\rho_{\xa j} \rho_{\xa \yb}} - \frac{1-\vij^2}{\rho_{\xa j}^2 \rho_{\xa \yb}}
  \right].
\end{align}
Using this result 
in Eq.~(\ref{eq:dcIntSectors}), we find 
\be\begin{split}
\overline{{\cal S}}_{\omega}[\Delta G_{ij}]_{\xa || \yb} & = 
-\frac{2\normA}{\ep} 
\int_0^{\infty} \frac{\dm \omega}{\omega^{1+2\ep}} \theta(1-\omega)
  \frac{ \omega  (2 + 3 \omega + 2 \omega^2)}{(1+\omega)^4}
  \\
 &  \times  \left \langle
 (\rho_{t\xa}^{4\ep}-1) 
 \left ( \frac{(2-\rho_{\xa j}) }{\rho_{\xa j} \rho_{\xa \yb}} - \frac{1-\vij^2}{\rho_{\xa j}^2 \rho_{\xa \yb}}
 \right )
 \right\rangle_{\xa\yb}.
\end{split}
\ee
We integrate over the energy $\omega$, and obtain 
\begin{equation}
  \begin{split}
    \gamma_\omega  =&  \int \limits_{0}^{1}  \frac{{\rm d} \omega}{\omega^{1+2\ep} }
                      \frac{\omega(2+3\omega+2\omega^2)}{(1+\omega)^4} =
                      \frac{11}{12}
                      + \ep \left(\frac{1}{12}+\frac{11}{3}\ln(2)\right)\\
                    & 
                      + \ep^2 \left(\frac{1}{3}+\frac{11}{18}\pi^2\right) 
                      + \ep^3 \left(\frac{4}{3}\ln(2) + 11 \zeta_3 \right)
                      +\mathcal{O}\left(\ep^4\right).
  \end{split}
\end{equation}
The remaining integration over directions 
of $\yb$ is straightforward. Putting everything together, we find
\be
\begin{split}
\overline{{\cal S}}_{\omega}[\Delta G_{ij}]_{\xa || \yb} & = 
\frac{(1-2\ep)\normA \gamma_\omega}{\ep^2}  
\left\langle
(\rho_{t\xa}^{4\ep}-1)  \; \left[
  \frac{(2-\rho_{\xa j}) }{\rho_{\xa j} } - \frac{1-\vij^2}{\rho_{\xa j}^2 }
  \right]
\right\rangle_{\xa}.
\end{split}
\label{eq4.48}
\ee
It follows from Eq.~(\ref{eq4.48}), 
 that $\overline{\cal S}_\omega[\Delta G_{ij}]_{\xa || \yb}$ has a $1/\ep$ singularity, since the integrand 
 is ${\cal O}(\ep)$.
 Finally, we  express Eq.~(\ref{eq4.48})  through integrals 
described in the appendix and find 
\begin{equation}
  \begin{split}
    \label{eq:dcIntResults}
      \overline {S}_\omega[\Delta G_{ij}] _{\xa || \yb}
            =
      \frac{1-2\ep}{\ep^2}{\cal N}_A \gamma_\omega
      &
      \Big\{
      2\AImm{-4\ep}{1}-2\AIm{1}[\rho_{jj}]-\AIm{-4\ep}[\rho_{jj}]+1
      \\
      & \quad
      -(1-\vij^2)\left[I^{(2)}_{-4\ep,2}-\AIm{2}[\rho_{jj}]\right]
      \Big\}.
  \end{split}
\end{equation}

\subsection{The quantity 
$\overline{\cal S}_\omega[\Delta G_{ij} ]
_{\rm fin,a_1}$
}
In this section, we compute 
the quantity $\overline{\cal S}_\omega[\Delta G_{ij} ]
_{\rm fin,a_1}$ defined in Eq.~(\ref{eq4.22}). In that equation, one can integrate over the gluon energy 
$\omega$. This gives 
\be
\overline{\cal S}_\omega[\Delta G_{ij} ]
_{\rm fin,a_1}
= -\frac{\normA}{4\ep^2}
\left\langle
  \left[\left(\frac{\rho_{t \xa}}{\rho_{t \yb}}\right)^{2\ep}-1\right] 
  \Delta_{ij}
\right\rangle_{\xa\yb}.
\enq
The function $\Delta_{ij}$ defined in 
Eq.~(\ref{eq4.20}) is anti-symmetric 
w.r.t. $\xa \leftrightarrow  \yb$ permutations. 
Using this, we find 
\bnq
\overline{\cal S}_\omega[\Delta G_{ij} ]
_{\rm fin,a_1}= 
\frac{\normA}{8\ep^2}
\left\langle
  \left[\left(\frac{\rho_{t \yb}}{\rho_{t \xa}}\right)^{2\ep}-\left(\frac{\rho_{t \xa}}{\rho_{t \yb}}\right)^{2\ep}\right] 
  \Delta_{ij}
\right\rangle_{\xa\yb}.
\enq
Since 
\bns
\left(\frac{\rho_{t \yb}}{\rho_{t \xa}}\right)^{2\ep}-\left(\frac{\rho_{t \xa}}{\rho_{t \yb}}\right)^{2\ep}
&=
4 \ep \ln \frac{\rho_{t \yb}}{\rho_{t \xa}} + {\cal O}(\ep^3),
\ens
and   ${\cal O}(\ep^3)$ contribution is irrelevant for computing  
$\overline{\cal S}_\omega[\Delta G_{ij} ]
_{\rm fin,a_1}$ through ${\cal O}(\ep^0)$, it follows that the following equations hold 
\be
\overline{\cal S}_\omega[\Delta G_{ij} ]
_{\rm fin,a_1}= 
\frac{\normA}{8\ep^2}
\left\langle
  4 \ep \ln \frac{\rho_{t \yb}}{\rho_{t \xa} } \; 
  \Delta_{ij}
\right\rangle_{\xa\yb}
=
\frac{\normA}{4\ep^2}
\left\langle
  \left[\rho_{t \xa}^{-2\ep} - \rho_{t \xa}^{2\ep}\right] 
  \Delta_{ij}
\right\rangle_{\xa\yb}+{\cal O}(\ep).
\ee
Using the explicit expression 
for $\Delta_{ij}$, it is straightforward to write the above expression in terms of integrals described in the appendix.  We find 
\be
\begin{split}
& \overline{\cal S}_\omega[\Delta G_{ij} ]
_{\rm fin,a_1} 
=
\frac{\normA}{4\ep^2}
\Bigg\{
\Big [
 I^{(2)}_{0,0,1,2\ep}
-I^{(2)}_{0,1,0,2\ep}
-I^{(2)}_{1,-1,1,2\ep}
+I^{(2)}_{1,1,-1,2\ep}
\\
&
+(1-\vij^2)
 I^{(2)}_{0,2,1,2\ep}
-I^{(2)}_{0,1,2,2\ep}
-I^{(2)}_{1,2,0,2\ep}
+I^{(2)}_{1,0,2,2\ep}
\Big ] - (2\ep \to -2\ep) 
\Bigg\},
\end{split}
\ee
where the change in the sign of 
$\ep$ only applies to the last index of integrals $I^{(2)}_{a,b,c,d,2\ep}$. 

\subsection{Calculation of $\overline{{\cal S}}_{\omega}[G^{(0)}_{ij}]$}
\label{ssect4.4}

The last quantity we require is $\overline{{\cal S}}_{\omega}[G^{(0)}_{ij}]$;
it is defined in 
Eq.~(\ref{eq4.8}). 
We remind the reader that this quantity 
can be understood 
as the integral of the eikonal  function in the  case where the rest frame of the parton $i$ and the laboratory frame coincide, and where the parton $j$ moves with the velocity $\vij$. 
To compute $\overline{{\cal S}}_{\omega}[G^{(0)}_{ij}]$, we
write it as an integral over the energy
$\omega$ 
\bnq
\overline{\cal S}_{\omega}[G^{(0)}_{ij}]
=
\frac{\normA}{\ep} 
\int_0^{1} \dm \omega \; 
\mathcal{A}_{ij}(\omega).
\enq

To obtain the function ${\cal A}_{ij}(\omega)$,  we integrate  over the angles of partons $\xa$ and $\yb$ 
at fixed $\omega$. 
The  function $\mathcal{A}_{ij}(\omega)$, defined as 
\bnq
\mathcal{A}_{ij}(\omega)
=
\left\langle
\omega^{-1-2\ep} \overline{S}_{\omega} \; 
\left[\omega^2 \widetilde{S}_{ij}(\xa, \yb)\right]
\right\rangle_{\xa\yb},
\enq
 can be written 
 as an integral over the 
gluon momenta with fixed energies 
\bnq
\mathcal{A}_{ij}(\omega)
=
\int \frac{[\dm l_{\xa}] [\dm l_{\yb}]}{\Nep^2} \;
\delta(1-l_{\xa} \cdot P)
\; \delta(\omega-l_{\yb} \cdot P) \; 
\omega^{-2}
\overline{S}_{\omega}
\left[\omega^2 \widetilde{S}_{ij}(l_{\xa}, l_{\yb})\right].
\enq
The quantity  $\Nep$ is defined in Eq.~(\ref{eq3.3}), and 
$P=(1,\vec 0)$. 
The  function ${\cal A}_{ij}(\omega)$ can be computed using 
reverse unitarity~\cite{Anastasiou:2002yz}
in a straightforward way. 
Using the 
integration-by-parts technology \cite{tkachov:1981wb,chetyrkin:1981qh}, we express ${\cal A}_{ij}(\omega)$
in terms of eight master integrals 
\begin{align}
J_1 &= \MIang{1} , & 
J_2 &= \MIang{\frac{1}{D_2}} , & 
J_3 &= \MIang{\frac{1}{D_3}} , & 
J_4 &= \MIang{\frac{1}{D_2 D_3}} , & \nonumber \\ 
J_{5} &= \MIang{\frac{1}{D_1 D_4}} , & 
J_{6} &= \MIang{\frac{1}{D_4^2}} , & 
J_{7} &= \MIang{\frac{1}{D_4}} , & 
J_{8} &= \MIang{\frac{1}{D_2 D_4}} . & 
\label{eq: mis_G0}
\end{align}
The four inverse propagators 
$D_{1,..,4} $ read 
\bnq
D_1 = l_{\xa} \cdot l_{\yb}, \;
D_2 = l_{\xa} \cdot p_j, \;
D_3 = l_{\yb} \cdot p_j, \;
D_4 = l_{\xa} \cdot p_j+l_{\yb} \cdot p_j,
\enq
and the integration measure is defined according to the following equation
\be
\Big \langle X \Big \rangle = 
\int \frac{[\dm l_{\xa}] [\dm l_{\yb}]}{
\Nep^2}
\delta(1-l_{\xa} \cdot P) \delta(\omega-l_{\yb} \cdot P)
\; X .
\enq
The four-momentum  of the parton $j$  is $p_j = 
(1, \vec v_{ij})$,  and $\overline {\cal S}_\omega[G_{ij}^{(0)}]$ depends on the absolute
value of the relative velocity $v_{ij}$.  It turns out to be  convenient to write the result in terms of the 
following variable
\be
\eta_{ij} = \frac{1-v_{ij}}{1+v_{ij}}.
\label{eq4.61}
\ee

We use the differential 
equations in $v_{ij}$ to compute the required integrals; the boundary conditions are  easily obtained 
by computing the relevant  integrals at $v_{ij} = 0$.  The calculation is described in 
 Appendix~\ref{app: g0_int}. 
 Using the results for the master integrals, we find 
\bns
& \overline{{\cal S}}_{\omega}[G^{(0)}_{ij}]
=
\frac{{\cal N}_A}{\ep^2} \Bigg [
-\frac{11}{6}-\frac{11 \ln (\eta_{ij} )}{12 \vij}
+
\ep \Bigg (
\frac{\ln^2(\eta_{ij} )}{2}+\frac{83}{9}-\frac{22 \ln (2)}{3}
+\frac{1}{\vij}
\left[
\frac{\ln^3(\eta_{ij} )}{24}
\right.
\\
& \left.
+\frac{2 \ln^2(\eta_{ij} )}{3}-\frac{8}{3} \text{Li}_2(1-\eta_{ij})
   -\frac{11}{3} \ln (2) \ln (\eta_{ij} )+\frac{131 \ln (\eta_{ij} )}{18}
\right] 
+\frac{1}{\vij^2}
\left[
-\frac{\ln ^3(\eta_{ij} )}{24} 
\right.
\\
& \left. -\frac{\ln ^2(\eta_{ij} )}{4}-\frac{\pi^2}{12}  \ln (\eta_{ij} )+\text{Li}_3(\eta_{ij}
   )
   -\frac{1}{2} \ln (\eta_{ij})  \text{Li}_2(\eta_{ij} )-\zeta_3
\right]
\Bigg )
+\mathcal{O}(\ep^2)
\Bigg ].
\label{eq4.62}
\ens
We do not show here the ${\cal O}(\ep^2)$ term 
in the square brackets, required to obtain 
$\overline{\cal S}_\omega[G_{ij}^{(0)}]$ through ${\cal O}(\ep^0)$.  The corresponding result can be found
in the 
ancillary file provided with this submission.  

\section{The final results: analysis, implementation and checks}
\label{sect:sec5}

The final result for the integral of the double-emission eikonal function for two soft gluons is obtained by combining different terms displayed in  Eq.~(\ref{eq4.23}). The strongly-order contribution ${\cal S}_\omega[G_{ij}]$ is further split into analytic and numerical pieces as shown in Eq.~(\ref{eq4.40}).  A similar computation has been  performed for the soft 
$q \bar q$ pair;  for brevity, we do not discuss it and proceed directly to the results.

\subsection{Analytic expressions of the poles}

In the preceding  section we have isolated all 
divergent contributions needed  for the computation of the integrated eikonal function. Hence, 
we are in the position to combine them, and obtain the divergent part of the integral  in an analytic form. 

Many integrals required for this computation are obtained by solving the   differential equations discussed in appendices; 
the results are naturally expressed in terms of generalized   polylogarithms (GPLs)~\cite{Goncharov:1998kja, Goncharov:2001iea}. 
Unfortunately,  arguments of these GPLs are very complicated, especially because multiple square roots of 
kinematic variables appear in the course of simplifying systems of  differential equations.  
This leads to significant problems when a numerical evaluation of such an analytic result  is attempted. 

To devise a representation 
of the  analytic result suitable for fast numerical evaluation, 
 we follow the same procedures as in 
Ref.~\cite{Horstmann:2025hjp}. To this end, we employ the  symbol technique~\cite{Goncharov:2010jf, Duhr:2011zq} and 
use it to express all  GPLs up to weight four through   logarithms, classical
polylogarithms $\textrm{Li}_n$ ($n=2,3,4$) and an additional function
$\textrm{Li}_{2,2}$, defined as
\begin{equation}
\label{eq:Li22def}
\mathrm{Li}_{2,2}(z_1,z_2) = \sum\limits_{i>j>0}^\infty \frac{z_1^i z_2^j}{i^2 j^2} = 
\sum\limits_{i=1}^\infty \sum\limits_{j=1}^\infty 
 \frac{z_1^i}{(i+j)^2} 
 \frac{(z_1 z_2 )^j}{j^2}.
\end{equation}
Since the problem of finding suitable candidate functions
to express  linear combinations of GPL's
in terms of polylogs and ${\rm Li}_{2,2}$ is equivalent to an optimization problem, 
we  used  the program \texttt{Gurobi}~\cite{gurobi} to improve the efficiency of this procedure. 
In the process of manipulating GPLs,
we heavily relied on the package \texttt{PolyLogTools}~\cite{Duhr:2019tlz,
Maitre:2005uu, Maitre:2007kp} and used the library
\texttt{GiNaC}~\cite{Bauer:2000cp, Vollinga:2004sn} to evaluate GPLs
numerically.
Finally, we make sure that all functions that appear 
in the final result are real-valued in the physical region.

Below  we showcase the divergent parts of the integrated double-emission eikonal functions
for both gluon and quark cases. We write 
\begin{align}
\frac{\intSS[\widetilde {\cal S}_{ij}]}{\normA} &= 
\frac{\overline{{\cal S}}_{\omega}[G^{(0)}_{ij}]}{\normA}  
-\frac{1}{\ep^3}\left[1+\frac{\ln(\eta_{ij})}{v_{ij}}\right]
+\frac{1}{\ep^2}
\Bigg\{
5-2 \ln\left(1-\beta_i^2\right) 
+\frac{\ln(\eta_i)}{\beta_i}+ \frac{\ln(\eta_j)}{\beta_j}
\nonumber \\
&+\frac{q_{3}}{\vij}
+\frac{1}{\vij^2}\frac{\ln^2(\eta_{ij})}{4}
\Bigg\}
+
\frac{1}{\ep}
\Bigg\{
-\frac{\ln \left(\eta _i\right) \ln \left(\eta _{ij}\right)}{2\beta _i v_{ij}}-\frac{\ln \left(\eta _{ij}\right) \ln \left(\eta _j\right) }{2v_{ij} \beta
   _j}
\label{eq4.64}
\\
  & +\frac{q_{2}}{\beta _i}+\frac{q_{7}}{v_{ij}}+\frac{q_5}{v_{ij}^2}+\frac{ q_{1}}{\beta _j}+q_{4}
\Bigg\}
+\mathcal{O}(\ep^0)
,
\nonumber 
\\
\frac{\intSS[\widetilde{\mathcal{I}}_{ij}]}{\normA} &= 
\frac{1}{\ep^2}\left[-\frac{\ln \left(\eta _{ij}\right)}{6 v_{ij}}-\frac{1}{3}\right]
\nonumber 
\\
& +\frac{1}{\ep}\Bigg\{
\frac{37}{18}-\frac{2}{3}\ln\left(1-\beta_i^2\right)
+\frac{\ln \left(\eta _i\right)}{3 \beta_i}
+\frac{\ln \left(\eta _j\right)}{3 \beta_j}
+
\frac{q_6}{\vij}
\Bigg\}
+\mathcal{O}(\ep^0),
\label{eq4.65}
\end{align}
where $\eta_{ij}$ is defined in Eq.~(\ref{eq4.61}) and 
\bnq
\eta_i = \frac{1-\beta_i}{1+\beta_i}, \quad
\eta_j = \frac{1-\beta_j}{1+\beta_j}.
\enq
The divergent part of the  quantity $\overline {\cal S}_\omega[G_{ij}^{(0)}]$ can be found in 
Eq.~(\ref{eq4.62}). 
The results shown in Eqs.~(\ref{eq4.64},\ref{eq4.65}) contain quantities $q_{1,\dots,7}$. They read 
\begin{align}
q_{1} &= 
   4 \text{Li}_2\left(1-\frac{\eta _i \eta
   _{ij}}{z^2}\right)-\text{Li}_2\left(1-\frac{\eta _i \eta
   _{ij}}{z}\right)-\text{Li}_2\left(1-\frac{\eta _i}{z}\right)
   -\text{Li}_2\left(1-\frac{\eta _{ij}}{z}\right)
   +\text{Li}_2(1-z)
   \nn &
   -\frac{3}{2} \ln ^2\left(\eta _i\right)-4 \ln (2) \ln \left(\eta _i\right)+4 \ln \left(\eta _i\right) \ln \left(\eta _i+1\right)-\frac{1}{2} \ln
   \left(\eta _i\right) \ln \left(\eta _{ij}\right)
   +4 \ln \left(\eta _i+1\right) \ln
   \left(\eta _{ij}\right)
   \nn &
   +\ln (z) \ln \left(\eta_i\right)
   -8 \ln (z) \ln \left(\eta _i+1\right)
   +\frac{1}{2} \ln ^2\left(\eta _{ij}\right)-4
   \ln (2) \ln \left(\eta _{ij}\right)
   -3 \ln (z) \ln
   \left(\eta _{ij}\right)+\frac{7 \ln ^2(z)}{2}
   \nn &
   +8 \ln (2) \ln (z)-\frac{19
   \ln (z)}{3}+\frac{19 \ln
   \left(\eta _i\right)}{6}+\frac{19 \ln \left(\eta_{ij}\right)}{6}
,\nn
q_{2} &= 
   \text{Li}_2\left(1-\frac{\eta _i \eta
   _{ij}}{z}\right)-4 \text{Li}_2\left(1-\eta _i\right)+\text{Li}_2\left(1-\frac{\eta
   _i}{z}\right)
   -\text{Li}_2\left(1-\frac{\eta
   _{ij}}{z}\right)+\text{Li}_2(1-z)
   \nn &
   +\frac{3}{2} \ln ^2\left(\eta _i\right)+4 \ln (2) \ln \left(\eta _i\right)-4 \ln \left(\eta _i\right) \ln \left(\eta _i+1\right)+\frac{1}{2} \ln
   \left(\eta _i\right) \ln \left(\eta _{ij}\right)
   +\frac{\ln ^2(z)}{2}
   \nn &
   -\ln (z) \ln \left(\eta _i\right)-\frac{19 \ln
   \left(\eta _i\right)}{6}
,\nn
q_{3} &= 
2 \text{Li}_2\left(1-\frac{\eta _i \eta _{ij}}{z}\right)
-2 \text{Li}_2\left(1-\frac{\eta _i}{z}\right)
+2 \text{Li}_2\left(1-\frac{\eta _{ij}}{z}\right)
+2 \text{Li}_2(1-z)
-3 \text{Li}_2\left(1-\eta _{ij}\right)
\nn
&+2 \ln \left(\eta _i+1\right) \ln \left(\eta _{ij}\right)
+\frac{1}{4} \ln
   ^2\left(\eta _{ij}\right)
-2 \ln (2) \ln \left(\eta _{ij}\right)
-2 \ln (z) \ln \left(\eta
   _{ij}\right)
+\ln ^2(z)
\nn
&+2\ln\left(\eta _{ij}\right)
,\nn
q_{4} &= -3 \ln ^2\left(\eta _i\right)-8 \ln ^2\left(\eta _i+1\right)-8 \ln (2) \ln \left(\eta
   _i\right)+8 \ln \left(\eta _i\right) \ln \left(\eta
   _i+1\right)
   +16 \ln (2) \ln \left(\eta _i+1\right)
   \nn &
   -\ln
   \left(\eta _i\right) \ln \left(\eta _{ij}\right)
   +2 \ln (z) \ln \left(\eta
   _i\right)-\frac{1}{4} \ln ^2\left(\eta _{ij}\right)
   +2 \ln (z) \ln \left(\eta
   _{ij}\right)-2 \ln ^2(z)-8 \ln ^2(2)
   \nn &
   +\frac{38 \ln (2)}{3}
   -\frac{38}{3} \ln \left(\eta _i+1\right)+\frac{19 \ln \left(\eta _i\right)}{3}
   -\frac{16}{3}
,\nn
q_5 &= 
    \ln \left(\eta
   _{ij}\right) \text{Li}_2\left(1-\frac{\eta _i}{z}\right)-\ln \left(\eta _{ij}\right)
   \text{Li}_2\left(1-\frac{\eta _i \eta _{ij}}{z}\right)
   +\ln \left(\eta _{ij}\right) \text{Li}_2\left(1-\eta
   _{ij}\right)
   -\text{Li}_2(1-z) \ln \left(\eta _{ij}\right)
   \nn &
   -\ln \left(\eta
   _{ij}\right) \text{Li}_2\left(1-\frac{\eta _{ij}}{z}\right)
   -\frac{1}{4} \ln ^3\left(\eta
   _{ij}\right)+\ln (2) \ln ^2\left(\eta _{ij}\right)
   +\ln (z) \ln
   ^2\left(\eta _{ij}\right)-\frac{1}{2} \ln ^2(z) \ln \left(\eta _{ij}\right)
   \nn &
   -\ln \left(\eta _i+1\right) \ln ^2\left(\eta _{ij}\right)
   -\frac{3}{4} \ln ^2\left(\eta_{ij}\right)
,\nn
q_{6} &= 
\frac{2}{3}
   \text{Li}_2\left(1-\frac{\eta _i \eta _{ij}}{z}\right)-\frac{2}{3}
   \text{Li}_2\left(1-\frac{\eta _i}{z}\right)
   -\frac{2 \text{Li}_2\left(1-\eta
   _{ij}\right)}{3}+\frac{2}{3} \text{Li}_2\left(1-\frac{\eta
   _{ij}}{z}\right)+\frac{2 \text{Li}_2(1-z)}{3}
   \nn &
   +\frac{1}{6} \ln ^2\left(\eta_{ij}\right)
   -\frac{4}{3} \ln (2) \ln \left(\eta _{ij}\right)-\frac{2}{3} \ln (z) \ln \left(\eta _{ij}\right)
   +\frac{2}{3} \ln \left(\eta _i+1\right) \ln \left(\eta _{ij}\right)
   \nn &
   +\frac{\ln ^2(z)}{3}
   +\frac{37 \ln \left(\eta _{ij}\right)}{36}
   ,\nn
q_{7} &= 
6 \text{Li}_3\left(\frac{\eta _i-z}{\eta _i \left(1-\eta _{ij}\right)}\right)-6
   \text{Li}_3\left(\frac{\left(\eta _i-z\right) \eta _{ij}}{z \left(1-\eta
   _{ij}\right)}\right)-4 \text{Li}_3\left(-\frac{z \left(\eta _i-z\right)}{\eta _i \left(\eta
   _{ij}-z\right)}\right)
   \nn &
   +4 \text{Li}_3\left(-\frac{\left(\eta _i-z\right) \eta _{ij}}{z
   \left(\eta _{ij}-z\right)}\right)+2 \text{Li}_3\left(-\frac{\eta _i \eta
   _{ij}-z}{z}\right)-4 \text{Li}_3\left(-\frac{\eta _i \eta _{ij}-z}{(1-z) z}\right)
   \nn & 
   +6
   \text{Li}_3\left(-\frac{\eta _i \eta _{ij}-z}{z \left(1-\eta _{ij}\right)}\right)-6
   \text{Li}_3\left(-\frac{\eta _i \eta _{ij}-z}{\eta _i \left(1-\eta
   _{ij}\right)}\right)-2 \text{Li}_3\left(\frac{\eta _i \eta _{ij}-z}{\eta _i \eta
   _{ij}}\right)
   \nn & 
   +4 \text{Li}_3\left(-\frac{z \left(\eta _i \eta _{ij}-z\right)}{(1-z)
   \eta _i \eta _{ij}}\right)-4 \text{Li}_3\left(\frac{\eta _i \eta _{ij}-z}{\eta
   _{ij}-z}\right)+4 \text{Li}_3\left(\frac{\eta _i \eta _{ij}-z}{\eta _i \left(\eta
   _{ij}-z\right)}\right)
   \nn &
   +4 \text{Li}_3\left(\frac{\eta _i-z}{1-z}\right)-2
   \text{Li}_3\left(-\frac{\eta _i-z}{z}\right)+2 \text{Li}_3\left(\frac{\eta _i-z}{\eta _i}\right)-4
   \text{Li}_3\left(\frac{\eta _i-z}{(1-z) \eta _i}\right)
   \nn &
   -3 \text{Li}_3\left(1-\eta
   _{ij}\right)+3 \text{Li}_3\left(-\frac{1-\eta _{ij}}{\eta _{ij}}\right)+6
   \text{Li}_3\left(\frac{1-\eta _{ij}}{1-z}\right)-6 \text{Li}_3\left(\frac{z \left(1-\eta
   _{ij}\right)}{(1-z) \eta _{ij}}\right)
   \nn &
   -6 \text{Li}_3\left(-\frac{1-\eta
   _{ij}}{\eta _{ij}-z}\right)+6 \text{Li}_3\left(-\frac{z \left(1-\eta
   _{ij}\right)}{\eta _{ij}-z}\right)+2 \text{Li}_3\left(-\frac{\eta
   _{ij}-z}{z}\right)-2 \text{Li}_3\left(\frac{\eta _{ij}-z}{\eta _{ij}}\right)
   \nn &
   +2  \text{Li}_3(1-z)-2 \text{Li}_3\left(-\frac{1-z}{z}\right)
   +4 \ln \left(\eta _i\right) \text{Li}_2(1-z)-8 \ln \left(\eta _i+1\right) \text{Li}_2(1-z)
   \nn &
   +8 \ln (2)
   \text{Li}_2(1-z)+8 \ln \left(\eta _{ij}\right)
   \text{Li}_2\left(-\frac{1-\eta _i}{2 \eta _i}\right)+8 \ln \left(\eta _{ij}\right)
   \text{Li}_2\left(\frac{1-\eta _i}{\eta _i+1}\right)
   \nn &
   -7 \ln \left(\eta _i\right)
   \text{Li}_2\left(1-\frac{\eta _i}{z}\right)+8 \ln \left(\eta _i+1\right)
   \text{Li}_2\left(1-\frac{\eta _i}{z}\right)-8 \ln (2) \text{Li}_2\left(1-\frac{\eta
   _i}{z}\right)
   \nn &
   -3 \ln (z)
   \text{Li}_2\left(\frac{\eta _i-z}{\eta _i \left(1-\eta _{ij}\right)}\right)+3 \ln
   \left(\eta _i\right) \text{Li}_2\left(\frac{\eta _i-z}{\eta _i \left(1-\eta
   _{ij}\right)}\right)+6 \ln \left(\eta _{ij}\right) \text{Li}_2\left(\frac{\eta
   _i-z}{\eta _i \left(1-\eta _{ij}\right)}\right)
   \nn &
   -9 \ln \left(\eta _i\right)
   \text{Li}_2\left(1-\eta _{ij}\right)+12 \ln \left(\eta _i+1\right) \text{Li}_2\left(1-\eta
   _{ij}\right)-12 \ln (2) \text{Li}_2\left(1-\eta _{ij}\right)
   \nn &
   +5 \ln \left(\eta _i\right)
   \text{Li}_2\left(\frac{1-\eta _{ij}}{1-z}\right)+6 \ln (z) \text{Li}_2\left(\frac{z
   \left(1-\eta _{ij}\right)}{(1-z) \eta _{ij}}\right)-\ln \left(\eta _i\right)
   \text{Li}_2\left(\frac{z \left(1-\eta _{ij}\right)}{(1-z) \eta _{ij}}\right)
   \nn &
   -6 \ln
   \left(\eta _{ij}\right) \text{Li}_2\left(\frac{z \left(1-\eta _{ij}\right)}{(1-z)
   \eta _{ij}}\right)-3 \ln (z) \text{Li}_2\left(\frac{\left(\eta _i-z\right) \eta
   _{ij}}{z \left(1-\eta _{ij}\right)}\right)+3 \ln \left(\eta _i\right)
   \text{Li}_2\left(\frac{\left(\eta _i-z\right) \eta _{ij}}{z \left(1-\eta
   _{ij}\right)}\right)
   \nn &
   +4 \ln (z) \text{Li}_2\left(-\frac{z \left(\eta _i-z\right)}{\eta _i
   \left(\eta _{ij}-z\right)}\right)-5 \ln \left(\eta _{ij}\right)
   \text{Li}_2\left(-\frac{z \left(\eta _i-z\right)}{\eta _i \left(\eta
   _{ij}-z\right)}\right)-6 \ln (z) \text{Li}_2\left(-\frac{1-\eta _{ij}}{\eta
   _{ij}-z}\right)
   \nn &
   +5 \ln \left(\eta _i\right) \text{Li}_2\left(-\frac{1-\eta
   _{ij}}{\eta _{ij}-z}\right)-\ln \left(\eta _i\right) \text{Li}_2\left(-\frac{z
   \left(1-\eta _{ij}\right)}{\eta _{ij}-z}\right)+4 \ln (z)
   \text{Li}_2\left(-\frac{\left(\eta _i-z\right) \eta _{ij}}{z \left(\eta
   _{ij}-z\right)}\right)
   \nn &
   +\ln \left(\eta _{ij}\right)
   \text{Li}_2\left(-\frac{\left(\eta _i-z\right) \eta _{ij}}{z \left(\eta
   _{ij}-z\right)}\right)+4 \ln \left(\eta _i\right) \text{Li}_2\left(1-\frac{\eta
   _{ij}}{z}\right)-8 \ln \left(\eta _i+1\right) \text{Li}_2\left(1-\frac{\eta
   _{ij}}{z}\right)
   \nn &
   +8 \ln (2) \text{Li}_2\left(1-\frac{\eta
   _{ij}}{z}\right)+7 \ln
   \left(\eta _i\right) \text{Li}_2\left(\frac{\eta _i-z}{\eta _i \eta _{ij}-z}\right)+\ln
   \left(\eta _i\right) \text{Li}_2\left(\frac{\left(\eta _i-z\right) \eta _{ij}}{\eta _i \eta
   _{ij}-z}\right)
   \nn &
   -4 \ln \left(\eta _i\right) \text{Li}_2\left(\frac{(1-z) \left(\eta
   _i-z\right) \eta _{ij}}{\left(\eta _{ij}-z\right) \left(\eta _i \eta
   _{ij}-z\right)}\right)-4 \ln \left(\eta _i\right) \text{Li}_2\left(\frac{\left(\eta
   _i-z\right) \left(\eta _{ij}-z\right)}{(1-z) \left(\eta _i \eta
   _{ij}-z\right)}\right)
   \nn &
   +7 \ln \left(\eta _i\right) \text{Li}_2\left(1-\frac{\eta _i \eta
   _{ij}}{z}\right)-8 \ln \left(\eta _i+1\right) \text{Li}_2\left(1-\frac{\eta _i \eta
   _{ij}}{z}\right)+8 \ln (2) \text{Li}_2\left(1-\frac{\eta _i \eta
   _{ij}}{z}\right)
   \nn &
   -4 \ln (z) \text{Li}_2\left(-\frac{\eta _i \eta _{ij}-z}{(1-z)
   z}\right)-\ln \left(\eta _{ij}\right) \text{Li}_2\left(-\frac{\eta _i \eta
   _{ij}-z}{(1-z) z}\right)+3 \ln (z) \text{Li}_2\left(-\frac{\eta _i \eta _{ij}-z}{z
   \left(1-\eta _{ij}\right)}\right)
   \nn &
   -3 \ln \left(\eta _i\right) \text{Li}_2\left(-\frac{\eta
   _i \eta _{ij}-z}{z \left(1-\eta _{ij}\right)}\right)+3 \ln (z)
   \text{Li}_2\left(-\frac{\eta _i \eta _{ij}-z}{\eta _i \left(1-\eta
   _{ij}\right)}\right)
   -3 \ln \left(\eta _i\right) \text{Li}_2\left(-\frac{\eta _i \eta
   _{ij}-z}{\eta _i \left(1-\eta _{ij}\right)}\right)
   \nn &
   -4 \ln (z)
   \text{Li}_2\left(-\frac{z \left(\eta _i \eta _{ij}-z\right)}{(1-z) \eta _i \eta
   _{ij}}\right)+5 \ln \left(\eta _{ij}\right) \text{Li}_2\left(-\frac{z \left(\eta _i
   \eta _{ij}-z\right)}{(1-z) \eta _i \eta _{ij}}\right)
   +\frac{7}{2} \ln \left(\eta _i\right) \ln ^2\left(\eta _{ij}\right)
   \nn &
   -\ln \left(\eta
   _i+1\right) \ln ^2\left(\eta _{ij}\right)+6 \ln ^2\left(\eta _i\right) \ln \left(\eta
   _{ij}\right)+8 \ln (2) \ln \left(\eta _i\right) \ln \left(\eta _{ij}\right)
   \nn &
   -8 \ln \left(\eta _i\right) \ln
   \left(\eta _i+1\right) \ln \left(\eta _{ij}\right)
   -3 \ln ^2\left(\eta _{ij}\right) \ln \left(z-\eta _i
   \eta _{ij}\right)-8 \ln (z) \ln \left(\eta _i\right) \ln \left(\eta _{ij}\right)
   \nn &
   +8 \ln (z) \ln \left(\eta _i+1\right) \ln \left(\eta _{ij}\right)+3 \ln (z) \ln \left(\eta
   _{ij}\right) \ln \left(z-\eta _i \eta _{ij}\right)+2 \ln ^2(z) \ln \left(\eta
   _i\right)
   \nn &
   -4 \ln ^2(z) \ln \left(\eta _i+1\right)-\frac{1}{12} \ln ^3\left(\eta
   _{ij}\right)+3 \ln \left(1-\eta _{ij}\right) \ln ^2\left(\eta
   _{ij}\right)+\ln (2) \ln ^2\left(\eta _{ij}\right)
   \nn &
   -8 \ln (2) \ln (z) \ln \left(\eta_{ij}\right)
   -3 \ln (z) \ln
   \left(1-\eta _{ij}\right) \ln \left(\eta _{ij}\right)
   +4 \ln (2) \ln ^2(z)
   \nn &
   -\frac{7 \text{Li}_2(1-z)}{3}+\frac{7}{3} \text{Li}_2\left(1-\frac{\eta _i}{z}\right)
   +\frac{2
   \text{Li}_2\left(1-\eta _{ij}\right)}{3}-\frac{7}{3} \text{Li}_2\left(1-\frac{\eta _{ij}}{z}\right)
   -\frac{7}{3} \text{Li}_2\left(1-\frac{\eta _i \eta
   _{ij}}{z}\right)
   \nn &
   -\ln ^2\left(\eta
   _{ij}\right)+\frac{13}{3} \ln (2) \ln \left(\eta _{ij}\right)
   +\frac{7}{3} \ln (z) \ln \left(\eta _{ij}\right)-\frac{7
   \ln ^2(z)}{6}
   +\ln\left(\eta _i\right) \ln \left(\eta _{ij}\right)
   \nn &
   -\frac{13}{3} \ln \left(\eta _i+1\right) \ln \left(\eta _{ij}\right)
   -\frac{23 \ln
   \left(\eta _{ij}\right)}{6},
\end{align}
where $
z = \left ( \eta_i \; \eta_j \; \eta_{ij}
\right )^{1/2}$.

\subsection{Implementation, checks and numerical results}
\label{sec:code_checks}

\begin{table}[t]
  \centering
  \addtolength{\leftskip} {-2cm}
  \addtolength{\rightskip}{-2cm}
  \resizebox{1.1\textwidth}{!}{
  \noindent\begin{tabular}{|c|c||l|l||l|l|}
    \hline
    $N_{i,j}$ & $N_{\theta}$ & \multicolumn{1}{c|}{Gluons(MC)} & \multicolumn{1}{c||}{Gluons(DE)} & \multicolumn{1}{c|}{Quarks(MC)} & \multicolumn{1}{c|}{Quarks(DE)}\\
    \hline\hline
    2  & 2  &	$\mathtt{7.9436(20)	 \cdot 10^{-4}}$ & $\mathtt{7.9453708 \cdot 10^{-4}}$ & $\mathtt{-1.8834(5)  \cdot 10^{-6}}$	& $\mathtt{-1.8822142 \cdot 10^{-6}}$ \\
    2  & 9  &	$\mathtt{1.447303(9)\cdot 10^{-2}}$	& $\mathtt{1.4473147 \cdot 10^{-2}}$ &	$\mathtt{-3.6194(8)  \cdot 10^{-5}}$ & $\mathtt{-3.6179578 \cdot 10^{-5}}$\\
    2  & 16 &	$\mathtt{3.574673(24)\cdot 10^{-2}}$ & $\mathtt{3.5746842 \cdot 10^{-2}}$ & $\mathtt{-9.6734(18) \cdot 10^{-5}}$	& $\mathtt{-9.6709404 \cdot 10^{-5}}$\\
    2  & 23 &	$\mathtt{4.918967(27)\cdot 10^{-2}}$ & $\mathtt{4.9189925 \cdot 10^{-2}}$ &	$\mathtt{-1.39541(23) \cdot 10^{-4}}$	& $\mathtt{-1.3953750 \cdot 10^{-4}}$\\
    \hline
    9  & 2  &	$\mathtt{2.16337(10)\cdot 10^{-2}}$	& $\mathtt{2.1633179 \cdot 10^{-2}}$ &	$\mathtt{\phantom{-}1.6820(8)\cdot 10^{-4}}$ & $\mathtt{\phantom{-}1.6824617 \cdot 10^{-4}}$\\
    9  & 9  &	$\mathtt{3.67465(11)\cdot 10^{-1}}$	& $\mathtt{3.6747979 \cdot 10^{-1}}$ &	$\mathtt{\phantom{-}1.9800(15)\cdot 10^{-3}}$	    & $\mathtt{\phantom{-}1.9813388 \cdot 10^{-3}}$\\
    9  & 16 &	$\mathtt{8.16731(25)\cdot 10^{-1}}$	& $\mathtt{8.1673711 \cdot 10^{-1}}$ &	$\mathtt{\phantom{-}1.228(4)\cdot 10^{-3}}$	    & $\mathtt{\phantom{-}1.2358643 \cdot 10^{-3}}$\\
    9  & 23 &	$\mathtt{1.053284(32)	\phantom{\cdot 10^{-0}}}$	& $\mathtt{1.0532756}$  & $\mathtt{-1.106(5)\cdot 10^{-3}}$	    & $\mathtt{-1.1036785 \cdot 10^{-3}}$\\
    \hline
    16 & 2  &	$\mathtt{1.53621(8)\cdot 10^{-1}}$	& $\mathtt{1.5363543 \cdot 10^{-1}}$ &	$\mathtt{\phantom{-}4.4056(13)\cdot 10^{-3}}$	& $\mathtt{\phantom{-}4.4075784 \cdot 10^{-3}}$\\
    16 & 9  &	$\mathtt{2.10969(12)\phantom{\cdot 10^{-0}}}$	& $\mathtt{2.1097537}$ &	$\mathtt{\phantom{-}4.8217(20)\cdot 10^{-2}}$  & $\mathtt{\phantom{-}4.8238002 \cdot 10^{-2}}$\\
    16 & 16 &	$\mathtt{3.57157(33)\phantom{\cdot 10^{-0}}}$	& $\mathtt{3.5712515}$ &	$\mathtt{\phantom{-}4.554(5)\cdot 10^{-2}}$  & $\mathtt{\phantom{-}4.5557310 \cdot 10^{-2}}$\\
    16 & 23 &	$\mathtt{3.9824(5)\phantom{\cdot 10^{-0}}}$	& $\mathtt{3.9818443}$ &	$\mathtt{\phantom{-}2.278(9)\cdot 10^{-2}}$  & $\mathtt{\phantom{-}2.2664599 \cdot 10^{-2}}$\\
    \hline
    23 & 2  &	$\mathtt{3.55465(15)\phantom{\cdot 10^{-0}}}$	& $\mathtt{3.5546016}$ &	$\mathtt{\phantom{-}2.10520(26)\cdot 10^{-1}}$ & $\mathtt{\phantom{-}2.1051877 \cdot 10^{-1}}$\\
    23 & 9  &	$\mathtt{2.01624(27)\cdot 10^{1}}$	& $\mathtt{2.0161751 \cdot 10^{1}}$ &	$\mathtt{\phantom{-}1.1123(5) \phantom{\cdot}  \phantom{10^{-0}}}$& $\mathtt{\phantom{-}1.1119777}$\\
    23 & 16 &	$\mathtt{1.9866(6)\cdot 10^{1}}$	& $\mathtt{1.9860241 \cdot 10^{1}}$ &	$\mathtt{\phantom{-}9.204(12)\cdot 10^{-1}}$  & $\mathtt{\phantom{-}9.2047726 \cdot 10^{-1}}$\\
    23 & 23 &	$\mathtt{1.8636(9)\cdot 10^{1}}$	& $\mathtt{1.8630407 \cdot 10^{1}}$ &	$\mathtt{\phantom{-}7.514(15)\cdot 10^{-1}}$  & $\mathtt{\phantom{-}7.5231867 \cdot 10^{-1}}$\\
    \hline
  \end{tabular}
  }
  \caption{Comparison of the finite ${\cal O}(\ep^0)$ parts of the results for the integral of the double-emission eikonal for gluon and quark cases 
  for $\beta_i = \beta_j$.  Results obtained with the numerical code (MC) are compared with the calculation based on solving dedicated differential equations (DE) for master integrals in the $\beta_i = \beta_j$ case.}
  \label{tab:num-eq-be}
\end{table}

We combine the  results described  in the previous sections into a computer code where the analytic and numerical parts of the calculation  are put 
together. 
This code can be obtained 
from a {\sf git}-repository using the following command
\begin{center}
\texttt{git clone \href{https://github.com/apik/SSmm.git}{https://github.com/apik/SSmm.git}}
\end{center}
The code contains  routines for an  efficient numerical evaluation  of the polylogarithms $\textrm{Li}_n$,  and the 
$\textrm{Li}_{2,2}$ function, required for the analytical part of the result. These routines are constructed  following  
methods described in Ref.~\cite{frellesvig:2016ske}. Finite remainder functions are
integrated using  the Vegas Monte-Carlo method~\cite{Lepage:1980dq} as implemented in the \texttt{GSL} library\cite{galassi2002gnu}. 

We continue with the discussion of the checks of the calculation.  As we explained  in Section~\ref{sec:conv},
integrals of the eikonal functions depend on four  parameters -- $E_{\rm max}$, velocities 
of partons $i$ and $j$ that we refer to as  $\beta_{i,j}$, 
and the angle $\theta_{ij}$ between them. Apart from the dependence on $E_{\rm max}$, which is rather trivial, the dependence of the final results on the other three parameters is complex. 
We note that for the discussion in this section, we always show the results for the function 
\be
\label{eq5.1}
{\cal O}_\Xi(\beta_i,\beta_j,\cos \theta_{ij} ) = -\frac{ 4 E_{\rm max}^{4\ep}}{N_\ep^2} \intSS[\Xi_{ij}].
\ee
We also note that the numerical code 
in the {\sf git}-repository outputs  the results for this function as well. 

For the discussion of the numerical results, it is useful to consider a grid of  benchmark points. We  parametrize them by a  triplet of integer numbers 
$\{N_i,N_j,N_\theta\}$, which define $\beta_{i,j}$ and $\cos \theta_{ij}$ according to the following equation 
\begin{equation}
  \label{eq:N-grid-def}
  \beta_i = \frac{N_{i}}{25}, \quad  \beta_j = \frac{N_{j}}{25}, \quad   \cos{\theta_{ij}} = \cos{\left( \frac{N_{\theta} \pi}{25}  \right)},\quad N_x = 1,\dots, 24.
\end{equation}

A very useful feature of the integrated eikonal functions is  their  regularity in various kinematic limits, e.g. 
$\beta_{i,j} \to 0$ or $\beta_i \to \beta_j$.
We have used the latter limit 
to perform extensive checks of the results. 
Since this limit is regular, we directly  obtain the $\beta_i = \beta_j$ numerical values for the function ${\cal O}_\Xi$ from the computer code described  at the beginning of this section.
They are shown in Table~\ref{tab:num-eq-be} for several benchmark points.

Reference (DE) values in that table are  obtained from  a high-precision numerical
solution of the system of differential equations 
for the master integrals using  the package
\texttt{DiffExp}~\cite{Hidding:2020ytt}.
To explain this further, we note that if  $\beta_i = \beta_j=\beta$, 
the integrated eikonal function   depends on two variables only ($\beta$ and $\theta_{ij})$, so that methods similar to  the ones employed in 
our previous paper \cite{Horstmann:2025hjp} apply.
We then  perform the IBP
reduction for the equal-velocity case,  and construct a system of differential equations  for all required master integrals. Starting with simple boundary conditions at 
$\beta =  0$, 
where the dependence on $\theta_{ij}$ disappears, 
we  fix the value of the angle to the desired value and numerically solve  the system of equations with respect to a single variable $\beta$.  
We compare  the results of solving  the differential equations, and the results of the numerical integration 
in Table~\ref{tab:num-eq-be}.
For all considered kinematic points  excellent agreement is observed.   We note that an  exact  agreement is found  for the $1/\ep$  poles that we do not show in 
Table~\ref{tab:num-eq-be} for the sake of brevity. Finally, we note that we also compared the results of our calculation with the  analytic results of Ref.~\cite{Bizon:2020tzr}, that correspond to a particular case 
of $\beta_i = \beta_j$ and $\theta_{ij}  = \pi$, 
and found excellent agreement.

\begin{table}[t]
  \centering
  \begin{tabular}{|c|c|c||l|l||l|l|}
    \hline
    $N_{i}$ & $N_{j}$ & $N_{\theta}$
    & \multicolumn{1}{c|}{MC (gluons)} & \multicolumn{1}{c||}{SecDec(gluons)}
    & \multicolumn{1}{c|}{MC (quarks)} & \multicolumn{1}{c|}{SecDec(quarks)}\\\hline
    2 & 9  & 16
    & $\mathtt{0.3477025(18)}$	& $\mathtt{0.346(4)}$
    & $\mathtt{-1.5041(21)\cdot 10^{-4}}$& $\mathtt{-1.0(9)\cdot 10^{-4}}$
    \\
    9  & 16  & 23
    & $\mathtt{2.40990(8)}$	& $\mathtt{2.397(32)}$
    & $\mathtt{\phantom{-}1.0100(14)\cdot 10^{-2}}$ & $\mathtt{\phantom{-}1.023(34)\cdot 10^{-2}}$
    \\
    16  & 23  & 9
    & $\mathtt{9.6427(3)}$	& $\mathtt{9.46(28)}$
    & $\mathtt{\phantom{-}4.5287(6)\cdot 10^1}$ & $\mathtt{\phantom{-}4.54(2)\cdot 10^1}$
    \\
    \hline
  \end{tabular}
  \caption{ Comparison of the finite parts for the gluon and quark integrated double-emission 
    eikonal functions (MC) with the results of direct numerical integration(SecDec).}
  \label{tab:num-secdec}
\end{table}

To check the  obtained results for two \emph{different}  velocities, $\beta_i\neq \beta_j$, we performed  direct numerical 
integration of the eikonal functions in the laboratory frame using the sector-decomposition 
approach as implemented in the \texttt{pySecDec} package~\cite{Heinrich:2023til}. Since the structure of singularities for  $\beta_{i,j}\neq 1$ 
is simpler than in the case with massless emitters, 
it is  straightforward to subtract the collinear $\xa || \yb$ singularity, and then extract the soft singularity  if needed. 

To prepare a suitable input for \texttt{pySecDec}, we write the function 
${\cal O}_\Xi$ in the following way
\begin{equation}
\begin{split}
  \label{eq:SlabSD}
   {\cal O}_\Xi= \frac{1}{\ep} \int_0^1 \frac{\dm \omega} {\omega^{1+2\ep}}
 &  \left(
    \Bigl\langle (1-C_{\xa \yb}) \omega^2 \Xi_{ij}(\omega ,\vec{n}_\xa,\vec{n}_\yb)\Bigr\rangle_{\xa \yb }
    +  \Bigl\langle C_{\xa \yb} \omega^2 \Xi_{ij}(\omega,\vec{n}_\xa,\vec{n}_\yb)\Bigr\rangle_{\xa \yb}
  \right), 
  \end{split}
  \end{equation}
where the eikonal factors 
for gluons and quarks $\Xi_{ij} = \{\widetilde{\mathcal{S}}_{ij},
\widetilde{\mathcal{I}}_{ij}\}$  need to be
integrated over directions 
of partons $\xa$ and $\yb$, and the energy of the parton $\yb$.

The action
of the $C_{\xa \yb}$ operator in
Eq.~(\ref{eq:SlabSD}) makes
the first term finite 
in the $\xa || \yb$ limit, and simplifies the second one, rendering  it amenable to  analytic integration.
In particular, as we have seen earlier when  discussing 
the $C_{\xa \yb}$ limit, 
integrations over $\omega$ and the
directions  of partons $\xa$ and $\yb$ factorize.
We find 
\be
\begin{split}
  & \int_0^1 \frac{\dm \omega} {\omega^{1+2\ep}}\left(
    \Bigl\langle C_{\xa \yb }  \; \omega^2 \Xi_{ij}(\omega,\vec{n}_\xa,\vec{n}_\yb)\Bigr\rangle_{\xa,\yb}
  \right) 
  \\
  & =  W_\Xi\left(\rho_{ii}\AIm{2}[\rho_{ii}] + \rho_{jj}\AIm{2}[\rho_{jj}] - 2 \rho_{ij} \AImm{1}{1}[\rho_{ii},\rho_{jj},\rho_{ij}]\right),
\end{split}
\ee
with angular integrals provided in Appendices \ref{sec:angIntsDef} and~\ref{app: two_mass_int}. 
The functions $W_\Xi$ are  obtained by integrating 
over $\omega$; they are 
\begin{align}
  W_{\widetilde{\mathcal{S}}} & = - \frac{1 - 2\ep}{12 \ep^2} \left(
           6 - \ep (11 - \ep + 4\ep^2)- 2\ep^2 (11 + 4 \ep^2)(\psi_{1/2-\ep}-\psi_{-\ep})
           \right),\\
  W_{\widetilde{\mathcal{I}}} & = \frac{1 - 2\ep}{6\ep(1 - \ep)}\left(
           1 - 2\ep(1 - \ep) + \ep(2 - \ep(3 - 4\ep)(\psi_{1/2-\ep}-\psi_{-\ep})
           \right).
\end{align}

To integrate terms 
in  the integrand in Eq.~(\ref{eq:SlabSD}) with the 
$\xa || \yb$  divergences subtracted, we write them 
as follows
\begin{align}
  I^{\textrm{sd}}_{GG} & = \int_{0}^{1} {\rm d} \omega 
  \; \omega^{-1-2\ep}\left\langle \mathcal{F}_{GG}^{(0)}(\omega,\vec{n}_\xa,\vec{n}_\yb) + \ep \mathcal{F}_{GG}^{(1)}(\omega,\vec{n}_\xa,\vec{n}_\yb) \right\rangle_{\xa,\yb}, \\
  I^{\textrm{sd}}_{QQ} & = \int_0^{1} {\rm d} \omega  \; \omega^{-2\ep} \left\langle  \mathcal{F}_{QQ}^{(0)}(\omega,\vec{n}_\xa,\vec{n}_\yb)\right\rangle_{\xa,\yb}.
\end{align}
The functions $\mathcal{F}_{GG,QQ}$ are finite in the $\omega \to 0$ limit and are
independent of  the 
regularisation parameter $\ep$. Angular
integrations over the directions of 
the parton momenta have to be performed in $(d-1)$-dimensions.  We treat these integrations, as well as integrations over $\omega$ 
using the sector decomposition~\cite{Heinrich:2023til}.\footnote{We note that functions $\mathcal{F}_{GG,QQ}$ are non-singular, so it is possible and, in fact, beneficial to keep them \emph{implicit} during the sector-decomposition
process.
} 

The results of this evaluation  for  several benchmark points are shown 
in Table~\ref{tab:num-secdec}, where the finite 
parts of the integrated double-emission eikonal 
functions for 
three kinematic points are given.
We observe a rather satisfactory  agreement between the two independent calculations of   the integral of the double-emission eikonal function.    

We would like to conclude this section by showing  
the results for the integrated double-emission eikonal functions for a few benchmark points. These results, shown in 
Tables~\ref{tab:bench-gg},\ref{tab:bench-qq} can be used to verify that the numerical code from the 
{\sf git}-repository is interpreted and used properly, including the $\ep$-dependent normalization prefactor.

\begin{table}[t]
  \centering
  \resizebox{1.0\textwidth}{!}{
  \begin{tabular}{|ccc|llll|}
    \hline
    $\beta_i$ & $\beta_j$ & $\cos\theta_{ij}$ & \multicolumn{1}{c}{$\ep^{-3}$}   & \multicolumn{1}{c}{$\ep^{-2}$} & \multicolumn{1}{c}{$\ep^{-1}$} & \multicolumn{1}{c|}{$\ep^{0}$}\\
    \hline
      0.08 & 0.36 & -0.4257 & $\mathtt{5.7483627\cdot 10^{-2}}$  & $\mathtt{3.3156281 \cdot 10^{-2}}$ & $\mathtt{1.7083994\cdot 10^{-1}}$ & $\mathtt{3.477025(18)\cdot 10^{-1}}$\\
      0.36 & 0.64 & -0.9686 & $\mathtt{3.9188310\cdot 10^{-1}}$  & $\mathtt{2.9983506\cdot 10^{-1}}$ & $\mathtt{1.1508376}$ & $\mathtt{2.40990(8)}$\\
    0.64 & 0.92 & 0.4258 & $\mathtt{7.0543274\cdot 10^{-1}}$  & $\mathtt{1.3901524}$ & $\mathtt{4.1595510}$ & $\mathtt{9.6427(3)}$\\
    \hline
  \end{tabular}}
  \caption{ 
  Benchmark results 
  for the function ${\cal O}_{GG}$, c.f 
  Eq.~(\ref{eq5.1}).
  }
  \label{tab:bench-gg}
\end{table}

\begin{table}[t]
  \centering
  \resizebox{0.8\textwidth}{!}{
  \begin{tabular}{|ccc|lll|}
    \hline
    $\beta_i$ & $\beta_j$ & $\cos\theta_{ij}$ & \multicolumn{1}{c}{$\ep^{-2}$} & \multicolumn{1}{c}{$\ep^{-1}$} & \multicolumn{1}{c|}{$\ep^{0}$}\\
    \hline
    0.08 & 0.36 & -0.4257 & $\mathtt{1.9161209\cdot 10^{-2}}$ & $\mathtt{2.4858662\cdot 10^{-6}}$ & $\mathtt{-1.5041(21)\cdot 10^{-4}}$\\
    0.36 & 0.64 & -0.9686 & $\mathtt{1.3062770\cdot 10^{-1}}$ & $\mathtt{1.3410916\cdot10^{-2}}$& $\mathtt{\phantom{-}1.0100(14)\cdot 10^{-2}}$\\
    0.64 & 0.92 & 0.4258 & $\mathtt{2.3514424\cdot 10^{-1}}$ & $\mathtt{2.9111167\cdot 10^{-1}}$ & $\mathtt{\phantom{-}4.5287(6)\cdot 10^1}$\\
    \hline
  \end{tabular}}
  \caption{Benchmark results for the function ${\cal O}_{QQ}$, c.f. Eq.~(\ref{eq5.1}).}
  \label{tab:bench-qq}
\end{table}

\section{Conclusion}
\label{sec:concl}
In this paper, we computed the integrals of the double-emission eikonal functions for two massive emitters whose momenta are at an arbitrary angle to each other.  This result is one of a few ingredients required for  extending the nested soft-collinear subtraction scheme \cite{caola:2017dug} to processes with  massive partons.  

Our computation is based on the approach  already described in  Refs.~\cite{agarwal:2024gws,Horstmann:2025hjp} where  local subtractions were used to remove potential singularities from the eikonal functions. The subtracted terms are integrated analytically, whereas  finite remainders, suitable for numerical integration, are computed  in four space-time dimensions. 

We emphasize that all $1/\ep$ poles 
that appear in the integral of the double-soft eikonal function are calculated analytically. The analytic results 
are expressed in  terms of standard logarithmic and polylogarithmic functions up to weight four, as well as the function ${\rm Li}_{2,2}$. 
This simple representation enables fast and efficient evaluation  of the analytic part of the result. 

The numerical code  that combines the analytic and the  numerical  parts of the calculation  can be obtained from a {\sf git}-repository.  We have checked it  for various kinematic cases, coming to the conclusion that, with  very moderate runtimes, it  provides a per mille relative precision for generic velocities and scattering angles.

\section*{Acknowledgments}
This research was supported in part by  
Deutsche Forschungsgemeinschaft (DFG, German Research Foundation) under grant no.\ 396021762 - TRR 257.

\appendix

\section{Massless and single-massive angular integrals}
\label{sec:angIntsDef}

In this appendix, we define 
some of the  angular integrals that  are used for calculations in Section~\ref{sec:doubleemission}. 
These  integrals contain either massless propagators, or  a massive and a massless  propagator. 

We fix the normalization of angular integrals in such a way that the angular volume is equal to one, 
\bnq
\Big \langle 1 \Big \rangle_{\xa} = \Big \langle 1 \Big \rangle_{\xa \yb}  = 1.
\enq
We begin by introducing  the following angular integrals,
\begin{alignat}{3}
  \renewcommand\arraystretch{1.33}
  \label{eq:angIntsBrackets}
  & \Big \langle \frac{1}{\rho_{\xa x}^n} \Big \rangle_{\xa} && = \AIm{n}\left[\rho_{xx}\right],\;\;\;\; &&\rho_{xx} \neq 0,\\
  & \Big \langle \frac{1}{\rho_{\xa x}^a \rho_{\xa y}^b} \Big \rangle_{\xa} && = \AIzz{a}{b}{\rho_{xy}} , &&\rho_{xx} = \rho_{yy} = 0,
\end{alignat}
These integrals can be computed in a closed form, in terms of hypergeometric functions. The results  read
\begin{align}
  &\AIm{n}\left[ \rho_{11} \right]
   = \left( 1 + \sqrt{1-\rho_{11}} \right)^{-n}  {}_2F_1\left(n,1-\ep,2-2\ep; \frac{2 \sqrt{1-\rho_{11}}}{1+\sqrt{1-\rho_{11}}}  \right) \label{eq:angInt-1-m},\\
  &\AIzz{a}{b}{\rho_{12}}
   = \frac{\Gamma\left(2-2\ep\right)\Gamma\left(1-\ep -a \right)\Gamma\left(1-\ep -b \right)}{2^{a+b}\Gamma^2\left(1-\ep\right)\Gamma\left(2-2\ep - a -b \right)}
    {}_2F_1\left(a,b, 1-\ep; 1-\frac{\rho_{12}}{2} \right). \label{eq:angInt-2-00}
\end{align}

\section{Double-massive angular integrals}
\label{app: two_mass_int}

We consider angular integrals with two massive propagators  defined as
\bnq
\Big \langle \frac{1}{\rho_{\xa x}^{c_1} \rho_{\xa y}^{c_2}} \Big \rangle_{\xa}  = \AImm{c_1}{c_2}\left[ \rho_{xx}, \rho_{yy} , \rho_{xy}\right] ,\quad \rho_{xx} \neq 0, \rho_{yy} \neq 0, \rho_{xy} \neq 0.
\enq
They can be mapped onto loop integrals by using the reverse unitarity~\cite{Anastasiou:2002yz}
\bnq
I^{(2)}_{c_1,c_2} = \frac{\Nep}{(2\pi)^{d-1}} 
\int \dm^d{k} 
\frac{\delta^+(k^2) \, \delta(1-k \cdot P)}{\left(k \cdot l_1\right)^{c_1}\left(k \cdot l_2\right)^{c_2}},
\enq
where the scalar products among $P, l_1, l_2$ read
\bnq
P \cdot l_{1,2} = 1, \quad P^2 = 1, \quad
l_1^2 = \rho_{xx} = x, \quad
l_2^2 = \rho_{yy} = y, \quad
l_1 \cdot l_2 = \rho_{xy} = w.
\enq
The powers of the propagators $c_i$ may depend linearly on $\ep$. We write  $c_i = a_i + b_i \ep$ and assume that $a_i$ and $b_i$ are integers. 

We can reduce the number of integrals  with two massive propagators that  need to be computed, by using the integration-by-parts (IBP)  technology \cite{tkachov:1981wb,chetyrkin:1981qh}.  Because $c_i$ depends on $\ep$, one has to keep powers 
of propagators symbolic when solving the IBP identities.
We note that since the IBP identities relate integrals whose propagator powers differ by
integers,  integrals with different $b_i$'s cannot  be connected by IBP relations.

We use 
\texttt{Kira~3}~\cite{Lange:2025fba} 
to perform the  IBP reduction with symbolic propagator powers.
We find that all integrals $I_{c_1,c_2}^{(2)}$, with $c_i = a_i+\ep b_i$, 
$i=1,2$, can be expressed through  four master integrals 
\bnq
  \boldsymbol{f} = \left( 
I^{(2)}_{b_1 \ep, b_2 \ep}, \quad 
I^{(2)}_{1+b_1 \ep, b_2 \ep}, \quad 
I^{(2)}_{b_1 \ep, 1+b_2 \ep}, \quad
I^{(2)}_{1+b_1 \ep, 1+b_2 \ep} \right)^{\mathrm{T}}.
\enq
To compute them, we 
define a new integral basis 
\bnq
\boldsymbol{g} = \frac{1}{1-2\ep} \left( g_1, \quad g_2, \quad g_3, \quad g_4\right)^{\mathrm{T}},
\enq
where
\bns
g_1 &= 
\left[1-\left(b_1+b_2+2\right) \ep \right]I^{(2)}_{b_1 \ep, b_2 \ep} 
+ b_1 \ep I^{(2)}_{1+b_1 \ep, b_2 \ep} + b_2 \ep I^{(2)}_{b_1 \ep, 1+b_2 \ep}
, \\
g_2 &= \ep \sqrt{1-x} I^{(2)}_{1+b_1 \ep, b_2 \ep}, \\
g_3 &= \ep \sqrt{1-y} I^{(2)}_{b_1 \ep, 1+b_2 \ep}, \\
g_4 &= \ep \sqrt{w^2-x y} I^{(2)}_{1+b_1 \ep, 1+b_2 \ep}.
\ens
The integrals $g_{1,..,4}$ satisfy the differential equations of the form 
\bnq
\dm \boldsymbol{g} = \ep \left(\dm \mathbb{A}\right) \, \boldsymbol{g},
\label{eqb7}
\enq
where the matrix $\mathbb{A}$ reads
\bnq
\mathbb{A} = \sum_{n_1=0, n_2=0}^{n_1+n_2 \leq 2} b_1^{n_1} b_2^{n_2} \mathbb{A}_{n_1 n_2},
\enq
with
\bnq
\begin{split}
\mathbb{A}_{00} &=
\left(
\begin{array}{cccc}
 0 & 0 & 0 & 0 \\
 -\frac{L_7}{2} & L_3-L_1 & 0 & 0 \\
 -\frac{L_8}{2} & 0 & L_4-L_2 & 0 \\
 -\frac{L_9}{2} & -L_{11} & -L_{12} & L_6-L_5 \\
\end{array}
\right)
, \\
\mathbb{A}_{10} &=
\left(
\begin{array}{cccc}
 -\frac{L_1}{2} & -L_7 & 0 & 0 \\
 0 & -\frac{L_1}{2} & 0 & 0 \\
 0 & \frac{L_{10}}{2} & -\frac{L_2}{2}+L_4-\frac{L_6}{2} & \frac{L_{12}}{2} \\
 0 & -\frac{L_{11}}{2} & -\frac{L_{12}}{2} & \frac{L_2}{2}-L_5+\frac{L_6}{2} \\
\end{array}
\right)
, \\
\mathbb{A}_{01} &=
\left(
\begin{array}{cccc}
 -\frac{L_2}{2} & 0 & -L_8 & 0 \\
 0 & -\frac{L_1}{2}+L_3-\frac{L_6}{2} & \frac{L_{10}}{2} & \frac{L_{11}}{2} \\
 0 & 0 & -\frac{L_2}{2} & 0 \\
 0 & -\frac{L_{11}}{2} & -\frac{L_{12}}{2} & \frac{L_1}{2}-L_5+\frac{L_6}{2} \\
\end{array}
\right)
, \\
\mathbb{A}_{11} &=
\left(
\begin{array}{cccc}
 0 & -\frac{L_7}{2} & -\frac{L_8}{2} & \frac{L_9}{2} \\
 0 & 0 & 0 & 0 \\
 0 & 0 & 0 & 0 \\
 0 & 0 & 0 & 0 \\
\end{array}
\right)
, 
\mathbb{A}_{20} =
\left(
\begin{array}{cccc}
 0 & -\frac{L_7}{2} & 0 & 0 \\
 0 & 0 & 0 & 0 \\
 0 & 0 & 0 & 0 \\
 0 & 0 & 0 & 0 \\
\end{array}
\right)
, 
\mathbb{A}_{02} =
\left(
\begin{array}{cccc}
 0 & 0 & -\frac{L_8}{2} & 0 \\
 0 & 0 & 0 & 0 \\
 0 & 0 & 0 & 0 \\
 0 & 0 & 0 & 0 \\
\end{array}
\right).
\end{split}
\enq
In the above equations,  we have used the following short-hand notations
\begin{align}
  \label{eq: lettter}
  L_{1} &= \log (x),   & L_{2} &= \log (y),   &  \nonumber\\
  L_{3} &= \log (1-x), & L_{4} &= \log (1-y), & \nonumber\\
  L_{5} &= \log \left(w^2-x y\right), & L_{6} &= \log \left( (1-w)^2-(1-x)(1-y) \right), & \nonumber\\
  L_{7} &= \log \left(\frac{1-\sqrt{1-x}}{1+\sqrt{1-x}}\right) , & L_{8} &= \log  \left(\frac{1-\sqrt{1-y}}{1+\sqrt{1-y}}\right), & \\
  L_{9} &= \log \left(\frac{w-\sqrt{w^2-x y}}{w+\sqrt{w^2-x y}}\right), & L_{10} &= \log \left(\frac{1-w-\sqrt{1-x} \sqrt{1-y}}{1-w+\sqrt{1-x} \sqrt{1-y}}\right), &\nonumber\\
  L_{11} &= \log \left(\frac{w-x-\sqrt{1-x} \sqrt{w^2-x y}}{w-x+\sqrt{1-x} \sqrt{w^2-x y}}\right) , &  L_{12} &= \log \left(\frac{w-y-\sqrt{1-y} \sqrt{w^2-x y}}{w-y+\sqrt{1-y} \sqrt{w^2-x y}}\right). & \nonumber
\end{align}

To solve the differential equation 
(\ref{eqb7}), we use the boundary condition
\bnq
\boldsymbol{g}(x=1, y=1, w=1) = (1, 0, 0, 0)^{\mathrm{T}},
\enq
and rationalize the square roots in the differential equation, by changing variables 
\bnq
x = \frac{4 \eta _1}{\left(\eta _1+1\right){}^2}, \;
y = \frac{4 \eta _2}{\left(\eta _2+1\right){}^2}, \;
w = \frac{2 \left(\eta _1 \eta _2+z^2\right)}{\left(\eta _1+1\right) \left(\eta _2+1\right) z}.
\label{eq: rationalization}
\enq
The resulting differential equations are then simple enough, and the solution can be straightforwardly obtained in terms of  GPLs.

\subsection{Two-loop double-massive integrals}

In Section~\ref{ssect4.1}
we introduced the following integral family
\bnq
  I^{(2)}_{a_1, a_2, a_3, a_4+b_4\ep} 
  =
\left\langle 
  \frac{1}{
  \rho_{\xa \yb}^{a_1}
  \rho_{\xa y}^{a_2}
  \rho_{\yb y}^{a_3}
  \rho_{\xa x}^{a_4+b_4\ep}
  }
\right\rangle_{\xa \yb},
\enq
and used such  integrals to write the strongly-ordered contribution to the eikonal function 
${\cal S}_\omega[G_{ij}]$.
To compute these integrals we perform the IBP reduction with \texttt{Kira 3} and identify twelve master integrals 
\begin{align}
    f_{1} &= I^{(2)}_{0,0,0,b_4 \ep}, &
    f_{2} &= I^{(2)}_{0,0,0,1+b_4 \ep}, &
    f_{3} &= I^{(2)}_{0,1,0,b_4 \ep}, &
    f_{4} &= I^{(2)}_{0,1,0,1+b_4 \ep}, & \nn 
    f_{5} &= I^{(2)}_{0,0,1,b_4 \ep}, &
    f_{6} &= I^{(2)}_{0,0,1,1+b_4 \ep}, &
    f_{7} &= I^{(2)}_{0,1,1,b_4 \ep}, &
    f_{8} &= I^{(2)}_{0,1,1,1+b_4 \ep}, & \\
    f_{9} &= I^{(2)}_{1,0,1,b_4 \ep}, &
    f_{10} &= I^{(2)}_{1,-1,1,b_4 \ep}, &
    f_{11} &= I^{(2)}_{1,0,1,1+b_4 \ep}, &
    f_{12} &= I^{(2)}_{1,-1,1,1+b_4 \ep}. & \nonumber
\end{align}
We then construct  differential equations for these master integrals by differentiating them  with respect to   $x=\rho_{xx}$, $y=\rho_{yy}$ and $w = \rho_{xy}$,  and 
using the reduction to express the result in terms of $f_{1,2,..,12}$. 
It is also possible to construct a \emph{canonical} basis  
for these integrals that we will again refer to as 
$\boldsymbol{g}$.  
The relation between $\boldsymbol{g}$ and $\boldsymbol{f}$ reads 
\bnq
\boldsymbol{g} = \mathrm{T} \boldsymbol{f},
\enq
where
\bnq
\mathrm{T}
=
\left(
\begin{array}{ccc}
 T_{11} & 0 & 0 \\
 0 & T_{22} & 0 \\
 T_{31} & 0 & T_{33} \\
\end{array}
\right),
\enq
with
\bns
&
T_{11}
=
\left(
\begin{array}{cccc}
 \frac{\left(b_4+2\right) \ep -1}{2 \ep -1} & \frac{b_4 \ep }{1-2 \ep } & 0 & 0 \\
 0 & \frac{\sqrt{1-x} \ep }{1-2 \ep } & 0 & 0 \\
 0 & 0 & \frac{\sqrt{1-y} \ep }{1-2 \ep } & 0 \\
 0 & 0 & 0 & \frac{\ep  \sqrt{w^2-x y}}{1-2 \ep } \\
\end{array}
\right),\;
T_{31}
=
\left(
\begin{array}{cccc}
 0 & 0 & 0 & 0 \\
 0 & 0 & 0 & 0 \\
 0 & 0 & 0 & 0 \\
 \frac{\left(b_4+2\right) \left(b_4 \ep -1\right)}{2 b_4 (1-2 \ep)} & \frac{\left(b_4+2\right) \ep }{2(\ep -1)} & 0 & 0 \\
\end{array}
\right), \\
&
T_{33}
=
\left(
\begin{array}{cccc}
 0 & 0 & \frac{\ep ^2 \sqrt{w^2-x y}}{(1-2 \ep )^2} & 0 \\
 0 & 0 & 0 & \frac{\sqrt{1-x} \ep ^2}{(1-2 \ep )^2} \\
 \frac{\left(b_4+2\right) \sqrt{1-y} \ep ^2}{b_4 (1-2 \ep )^2} & 0 & 0 & 0 \\
 -\frac{2 \left(b_4+2\right) \ep ^2}{b_4 (1-2 \ep )^2} & \frac{\left(b_4+2\right) \ep  \left(\left(b_4+2\right) \ep -1\right)}{b_4 (1-2 \ep )^2} & 0 & -\frac{\left(b_4+2\right) \ep ^2}{(1-2 \ep )^2}
   \\
\end{array}
\right),
\ens
and
\bnq
T_{22} = \sqrt{1-y}\frac{\ep}{1-2\ep} T_{11}.
\enq
The corresponding matrix $\mathbb{A}$ in the $d\log$ form of differential equation ${\rm d} \boldsymbol{g} = 
\ep {\rm d} ( \mathbb{A} ) \; \boldsymbol{g}$ reads
\bnq
\mathbb{A} = \sum_{n=-1}^{2} b_4^{n} \mathbb{A}_{n},
\enq
with
\bnq
\mathbb{A}_{-1}=
\scalemath{0.9}{
\left(
\begin{array}{cccccccccccc}
 0 & 0 & 0 & 0 & 0 & 0 & 0 & 0 & 0 & 0 & 0 & 0 \\
 0 & 0 & 0 & 0 & 0 & 0 & 0 & 0 & 0 & 0 & 0 & 0 \\
 0 & 0 & 0 & 0 & 0 & 0 & 0 & 0 & 0 & 0 & 0 & 0 \\
 0 & 0 & 0 & 0 & 0 & 0 & 0 & 0 & 0 & 0 & 0 & 0 \\
 0 & 0 & 0 & 0 & 0 & 0 & 0 & 0 & 0 & 0 & 0 & 0 \\
 0 & 0 & 0 & 0 & 0 & 0 & 0 & 0 & 0 & 0 & 0 & 0 \\
 0 & 0 & 0 & 0 & 0 & 0 & 0 & 0 & 0 & 0 & 0 & 0 \\
 0 & 0 & 0 & 0 & 0 & 0 & 0 & 0 & 0 & 0 & 0 & 0 \\
 0 & 0 & 0 & 0 & -L_{12} & 0 & 0 & 0 & 0 & 0 & 0 & 0 \\
 0 & 0 & 0 & 0 & L_{10} & 0 & 0 & 0 & 0 & 0 & 0 & 0 \\
 \frac{L_8}{2} & 0 & 0 & 0 & 0 & 0 & 0 & 0 & 0 & 0 & 0 & 0 \\
 0 & 0 & 0 & 0 & 2 L_8 & 0 & 0 & 0 & 0 & 0 & 0 & 0 \\
\end{array}
\right)
},
\mathbb{A}_{2}=
\scalemath{0.9}{
\left(
\begin{array}{cccccccccccc}
 0 & -\frac{L_7}{2} & 0 & 0 & 0 & 0 & 0 & 0 & 0 & 0 & 0 & 0 \\
 0 & 0 & 0 & 0 & 0 & 0 & 0 & 0 & 0 & 0 & 0 & 0 \\
 0 & 0 & 0 & 0 & 0 & 0 & 0 & 0 & 0 & 0 & 0 & 0 \\
 0 & 0 & 0 & 0 & 0 & 0 & 0 & 0 & 0 & 0 & 0 & 0 \\
 0 & 0 & 0 & 0 & 0 & -\frac{L_7}{2} & 0 & 0 & 0 & 0 & 0 & 0 \\
 0 & 0 & 0 & 0 & 0 & 0 & 0 & 0 & 0 & 0 & 0 & 0 \\
 0 & 0 & 0 & 0 & 0 & 0 & 0 & 0 & 0 & 0 & 0 & 0 \\
 0 & 0 & 0 & 0 & 0 & 0 & 0 & 0 & 0 & 0 & 0 & 0 \\
 0 & 0 & 0 & 0 & 0 & 0 & 0 & 0 & 0 & 0 & 0 & 0 \\
 0 & 0 & 0 & 0 & 0 & 0 & 0 & 0 & 0 & 0 & 0 & 0 \\
 0 & 0 & 0 & 0 & 0 & 0 & 0 & 0 & 0 & 0 & 0 & 0 \\
 0 & \frac{L_7}{4} & 0 & 0 & 0 & 0 & 0 & 0 & 0 & \frac{L_7}{2} & 0 & 0 \\
\end{array}
\right)
},
\enq
and
\bnq
\mathbb{A}_{0}=
\scalemath{0.5}{
\begingroup 
\setlength\arraycolsep{-1pt}
\begin{pmatrix}
  0 & 0 & 0 & 0 & 0 & 0 & 0 & 0 & 0 & 0 & 0 & 0 \\
  -\frac{L_7}{2} & L_3-L_1 & 0 & 0 & 0 & 0 & 0 & 0 & 0 & 0 & 0 & 0 \\
  -\frac{L_8}{2} & 0 & L_4-L_2 & 0 & 0 & 0 & 0 & 0 & 0 & 0 & 0 & 0 \\
  -\frac{L_9}{2} & -L_{11} & -L_{12} & L_6-L_5 & 0 & 0 & 0 & 0 & 0 & 0 & 0 & 0 \\
  -\frac{L_8}{2} & 0 & 0 & 0 & L_4-L_2 & 0 & 0 & 0 & 0 & 0 & 0 & 0 \\
  0 & -\frac{L_8}{2} & 0 & 0 & -\frac{L_7}{2} & -L_1-L_2+L_3+L_4 & 0 & 0 & 0 & 0 & 0 & 0 \\
  0 & 0 & -\frac{L_8}{2} & 0 & -\frac{L_8}{2} & 0 & 2 L_4-2 L_2 & 0 & 0 & 0 & 0 & 0 \\
  0 & 0 & 0 & -\frac{L_8}{2} & -\frac{L_9}{2} & -L_{11} & -L_{12} & -L_2+L_4-L_5+L_6 & 0 & 0 & 0 & 0 \\
  \frac{L_9}{4} & \frac{L_{11}}{2} & 0 & 0 & 0 & 0 & 0 & 0 & L_1+L_2-3 L_5+2 L_6 & -L_{11} & -L_{12} &\frac{L_9}{2} \\
  \frac{L_7}{4} & \frac{L_1}{2}-\frac{L_2}{2}-L_3+\frac{L_6}{2} & 0 & 0 & 0 & -L_8 & 0 & 0 & L_{11} & L_3-L_1 & L_{10} & \frac{L_7}{2} \\
  \frac{L_8}{4} & 0 & 0 & 0 & 0 & 0 & 0 & 0 & L_{12} & L_{10} & L_4-L_2 & \frac{L_8}{2} \\
  0 & 0 & 0 & 0 & L_8 & -2 L_{10} & 0 & 0 & -2 L_9 & 2 L_7 & 2 L_8 & -L_1-L_2+L_6 \\
\end{pmatrix}
\endgroup
}
,
\enq
\bnq
\mathbb{A}_{1}=
\scalemath{0.6}{
\begingroup 
\setlength\arraycolsep{1pt}
\begin{pmatrix}
 -\frac{L_1}{2} & -L_7 & 0 & 0 & 0 & 0 & 0 & 0 & 0 & 0 & 0 & 0 \\
 0 & -\frac{L_1}{2} & 0 & 0 & 0 & 0 & 0 & 0 & 0 & 0 & 0 & 0 \\
 0 & \frac{L_{10}}{2} & -\frac{L_2}{2}+L_4-\frac{L_6}{2} & \frac{L_{12}}{2} & 0 & 0 & 0 & 0 & 0 & 0 & 0
   & 0 \\
 0 & -\frac{L_{11}}{2} & -\frac{L_{12}}{2} & \frac{L_2}{2}-L_5+\frac{L_6}{2} & 0 & 0 & 0 & 0 & 0 & 0 &
   0 & 0 \\
 0 & 0 & 0 & 0 & -\frac{L_1}{2} & -L_7 & 0 & 0 & 0 & 0 & 0 & 0 \\
 0 & 0 & 0 & 0 & 0 & -\frac{L_1}{2} & 0 & 0 & 0 & 0 & 0 & 0 \\
 0 & 0 & 0 & 0 & 0 & \frac{L_{10}}{2} & -\frac{L_2}{2}+L_4-\frac{L_6}{2} & \frac{L_{12}}{2} & 0 & 0 & 0
   & 0 \\
 0 & 0 & 0 & 0 & 0 & -\frac{L_{11}}{2} & -\frac{L_{12}}{2} & \frac{L_2}{2}-L_5+\frac{L_6}{2} & 0 & 0 &
   0 & 0 \\
 0 & 0 & 0 & 0 & 0 & 0 & 0 & 0 & \frac{L_2}{2}-L_5+\frac{L_6}{2} & -\frac{L_{11}}{2} &
   -\frac{L_{12}}{2} & 0 \\
 0 & 0 & 0 & 0 & 0 & 0 & 0 & 0 & 0 & -\frac{L_1}{2} & 0 & 0 \\
 0 & 0 & 0 & 0 & 0 & 0 & 0 & 0 & \frac{L_{12}}{2} & \frac{L_{10}}{2} & -\frac{L_2}{2}+L_4-\frac{L_6}{2}
   & 0 \\
 0 & \frac{L_7}{2} & 0 & 0 & 0 & -L_{10} & 0 & 0 & -L_9 & 2 L_7 & L_8 & -\frac{L_1}{2} \\
\end{pmatrix}
\endgroup
}.
\enq
The  logarithms $L_{1,..,12}$ that 
appear in the above equations are defined in Eq. \eqref{eq: lettter}.
The boundary conditions are easily fixed by using
\bnq
g_n(x=1, y=1, w=1) = \delta_{1n}.
\enq
Using the variable transformation as  in Eq. \eqref{eq: rationalization}, we remove square roots from the system of equations  and solve it 
in terms of GPLs.

\section{Integrals for  $\overline{\cal S}_{\omega}[G^{(0)}_{ij}]$}
\label{app: g0_int}

It remains to discuss the calculation of integrals that are needed for the term  $\overline{\cal S}_{\omega}[G^{(0)}_{ij}]$, c.f. 
Section~\ref{ssect4.4}.
The required eight  master integrals are displayed   in Eq.~\eqref{eq: mis_G0}.  We use them to define   a new basis
\begin{align}
\widetilde{J}_1 &= \frac{J_1}{\omega} ,& 
\widetilde{J}_2 &= \frac{\vij  J_2 \ep }{\omega (2 \ep -1)} ,& 
\widetilde{J}_3 &= \frac{\vij  J_3 \ep }{ (2 \ep -1)} ,& \nonumber \\
\widetilde{J}_4 &= \frac{\vij ^2 J_4 \ep ^2}{ (2 \ep -1)^2} ,& 
\widetilde{J}_{5} &= \frac{\vij  J_{5} (\omega+1) \ep ^2}{ (2 \ep -1)^2} ,& 
\widetilde{J}_{6} &= \frac{\vij ^2 J_{6} \ep}{ (2 \ep -1)^2} ,& \\
\widetilde{J}_{7} &= 
\rlap{$\displaystyle
\frac{\vij  J_{7} (1-4 \ep ) \ep }{(\omega-1) (2 \ep -1)^2}
+\frac{\vij  J_{6} (\omega+1) \ep }{(\omega-1) (2 \ep -1)^2}
+\frac{2 \vij  J_{5} (\omega+1) \ep ^2}{(\omega-1) (2 \ep -1)^2} , 
\qquad
\widetilde{J}_{8} = \frac{\vij ^2 J_{8} \ep ^2}{(2 \ep -1)^2} .
$} & \nonumber
\label{eq: utG0}
\end{align}
These integrals satisfy the following  canonical differential equations 
\bnq
d \boldsymbol{\widetilde{J}} = \ep (d \mathbb{A}) \boldsymbol{\widetilde{J}},
\enq
where
\bnq
\mathbb{A} =
\left(
\begin{array}{cc}
 \mathbb{A}_{11} & 0 \\
 \mathbb{A}_{21} & \mathbb{A}_{22}
\end{array}
\right),
\enq
with
\bns
\mathbb{A}_{11} &=
\scalemath{0.7}{
\left(
\begin{array}{cccc}
 -2 L_4 & 0 & 0 & 0 \\
 \frac{L_3}{2}-\frac{L_1}{2} & -L_1+2 L_2-L_3-2 L_4 & 0 & 0 \\
 \frac{L_3}{2}-\frac{L_1}{2} & 0 & -L_1+2 L_2-L_3-2 L_4 & 0 \\
 0 & \frac{L_3}{2}-\frac{L_1}{2} & \frac{L_3}{2}-\frac{L_1}{2} & -2 L_1+4 L_2-2 L_3-2 L_4 \\
\end{array}
\right)
},\\
\mathbb{A}_{21} &=
\scalemath{0.8}{
\left(
\begin{array}{cccc}
 \frac{L_1}{4}-\frac{L_3}{4} & 0 & 0 & 0 \\
 -\frac{L_1}{4}-\frac{L_3}{4}-\frac{L_5}{2}+\frac{L_6}{4}+\frac{L_7}{4} & 0 & 0 & 0 \\
 -\frac{L_1}{4}+\frac{L_3}{4}+\frac{L_6}{4}-\frac{L_7}{4} & 0 & 0 & 0 \\
 0 & \frac{L_3}{2}-\frac{L_1}{2} & 0 & 0 \\
\end{array}
\right)
}, \\
\mathbb{A}_{22} &=
\scalemath{0.6}{
\left(
\begin{array}{cccc}
 -L_1+2 L_2-L_3-2 L_5 & L_1-L_3 & L_4 & 0 \\
 L_1-L_3-L_6+L_7 & -L_1+4 L_2-L_3-2 L_5-L_6-L_7 & L_7-L_6 & 0 \\
 L_1+L_3+2 L_5-L_6-L_7 & -L_1+L_3-L_6+L_7 & 2 L_2-2 L_4-L_6-L_7 & 0 \\
 L_1-L_3 & L_4 & L_1-L_3 & -2 L_1+4 L_2-2 L_3-2 L_4 \\
\end{array}
\right)
}.
\ens
The logarithms $L_i$ in this case are given by
\begin{align*}
L_{1} &= \log (1-\vij), &
L_{5} &= \log (1+\omega), & \\
L_{2} &= \log (\vij), &
L_{6} &= \log (1+\omega+\vij-\omega\vij), & \\
L_{3} &= \log (1+\vij), &
L_{7} &= \log (1+\omega-\vij+\omega\vij). & \\
L_{4} &= \log (\omega), &
\end{align*}
The boundary conditions are easily fixed by using
\bnq
\widetilde{J}_n(\vij = 0) =  \omega^{-2\ep} \delta_{1n}.
\enq
Finding the solution  of the system of differential equations in terms of GPLs is straightforward. 

\bibliographystyle{JHEP}
\bibliography{SSmm}

\end{document}